\documentclass{aa}  

\usepackage{longtable}
\usepackage{pdflscape}
\usepackage{lscape}
\usepackage{graphicx}
\usepackage{xcolor}
\usepackage{txfonts}
\begin{document}

   \title{The effect of dynamical states on galaxy clusters populations}

   \subtitle{I. Classification of dynamical states}

   \author{S. V\'eliz Astudillo \inst{1} \and
   E. R. Carrasco \inst{2} \and
   J. L. Nilo Castell\'on \inst{1} \and
   A. Zenteno \inst{3} \and
   H. Cuevas \inst{1}}
   \institute{Departamento de Astronom\'ia, Universidad de La Serena, Av. Ra\'ul Bitr\'an 1305, La Serena, Chile \and
   International Gemini Observatory, NSF's NOIRLab, Casilla 603, La Serena, Chile \and
   Cerro Tololo Inter-American Observatory, NSF's NOIRLab, Casilla 603, La Serena, Chile}

   \date{Received September 15, 1996; accepted March 16, 1997}

 
  \abstract
   {Although the influence of galaxy clusters on galaxy evolution is relatively well understood, the impact of the dynamical states of these clusters is less clear. This series of papers explores how the dynamical state of galaxy clusters affects their galaxy populations' physical and morphological properties.}
   {The primary aim of this first paper is to evaluate the dynamical state of 87 massive ($M_{500} \geq 1.5 \times 10^{14} M_{\odot}$) galaxy clusters at low redshifts ($0.10 \leq z \leq 0.35$). This will allow us to have a well-characterized sample for analyzing physical and morphological properties in our next work.}
   {We employ six dynamical state proxies that utilize optical and X-ray imaging data. Principal Component Analysis is applied to integrate these proxies effectively, allowing robust classification of galaxy clusters into relaxed, intermediate, and disturbed states based on their dynamical characteristics.}
   {The methodology successfully segregates the clusters of galaxies into the three dynamical states. Examination of the projected galaxy distributions in optical wavelengths and gas distributions in X-ray further confirms the consistency of these classifications. The dynamical states of the clusters are statistically distinguishable, providing a clear categorization for further analysis.}
   {}


   \maketitle
%

\section{Introduction}

In the context of the standard cosmological model Lambda Cold Dark Matter ($\Lambda$CDM), small fluctuations in the initial density field are amplified by gravity, collapsing into dark matter halos, transitioning from a linear to a nonlinear regime \citep{Molnar16}. Following the hierarchical paradigm of structure formation, mergers with similarly sized systems and smooth accretions of smaller structures lead to the formation of galaxy clusters \citep{KravstovBorgani12}, which are the largest gravitationally bound systems in the Universe.

Due to their formation processes, galaxy clusters can exist in various dynamical states. These range from systems close to virialization to highly perturbed systems resulting from large-scale interactions. According to the $\Lambda$CDM cosmological model and observations, around 30\%-80\% of the clusters are in unrelaxed states, exhibiting notable substructures in optical and X-ray images \citep[e.g.,][]{DresslerShectman88,Santos+08,FakhouriMa10} and merger features in radio images \citep[e.g.,][]{Skillman+13,Golovich+19}

Galaxy clusters in a relaxed dynamical state are crucial for studying various astrophysical aspects. For instance, they allow us to investigate the advanced stages of structure evolution in the Universe, providing insight into how galaxy clusters have evolved from their initial states to their current form \citep[e.g.,][]{Voit+05}. Additionally, due to their higher degrees of symmetry, these systems are ideal for studying dark matter density profiles through gravitational lensing \citep[e.g.,][]{Umetsu+14}. Furthermore, since the hot gas forming the intracluster medium (ICM) tends to be in hydrostatic equilibrium in relaxed clusters, more precise studies of its temperature and density distribution can be conducted, offering valuable information about the physics of gas and its cooling history \citep[e.g.,][]{Vikhlinin+06}. Moreover, considering that relaxed clusters exhibiting strong gravitational lensing are more numerous than their disturbed counterparts, they are ideal systems for studying high-redshift galaxies that are magnified by this effect. \citep[e.g.,][]{Bayliss+11,Bouwens+14}.

On the other hand, unrelaxed clusters offer different aspects of the study. These systems provide insights into the processes of large-scale structure formation in the Universe, such as merger dynamics or shock waves and cold fronts \citep[e.g.,][]{Markevitch+02,Poole+06,Owers+14}. Additionally, interactions within galaxy clusters can displace dark matter and ICM components, offering an ideal scenario for studying both components and their interactions \citep[e.g.,][]{Clowe+06,Jee+14}. Moreover, galaxy cluster merger scenarios can be helpful in the field of particle physics to constrain the cross section of self-interacting dark matter particles \citep[e.g.,][]{Harvey+15} and in cosmology, providing essential tests for the standard $\Lambda$CDM cosmological model \citep[e.g.,][]{Thompson+15}.

Various proxies have been used to classify systems according to their dynamical states. For example, in the projected sky plane, the cluster morphology in radio images has been utilized \citep[e.g.,][]{Cassano+10}, as well as the galaxy density distribution \citep[e.g.,][]{Wen&Han15}, the surface brightness distribution in X-rays \citep[e.g.,][]{Jeltema+05,Nurgaliev+13}, and the offset between the brightest cluster galaxy (BCG) and the X-ray peak/centroid \citep[e.g.,][]{Mann&Ebeling12,Zenteno+20}. In addition, to identify cluster mergers along the line of sight, the shape of the velocity distribution has also been used as a proxy \citep[e.g.,][]{Ribeiro+13,deLosRios+16,carrasco2021}.

However, one of the most fundamental aspects of galaxy clusters is that they are ideal laboratories for studying galaxy evolution. These are the highest-density environments in the Universe's large-scale structure, where specific physical processes accelerate galaxy evolution, leading to morphological transformations, environmental quenching, and preferential galaxy distributions. Although differences in physical and morphological properties between clusters and the field or other low-density environments are well studied, the effect of the dynamical state of galaxy clusters on their member galaxies is not entirely clear. \citet{Ribeiro+13} found that the faint end of the luminosity function is more pronounced in relaxed clusters than in disturbed systems at low redshift ($z \leq 0.1$) in clusters with $M > 10^{14} M_{\odot}$. However, \citet{Zenteno+20} found results that were in disagreement. By studying a sample of 288 massive galaxy clusters ($M_{200} > 4 \times 10^{14}$ M$_{\odot}$) with multiwavelength information, they found that the luminosity function does not show statistically significant differences between relaxed and disturbed clusters at low redshift ($z \lesssim 0.55$). However, at a higher redshift ($z \gtrsim 0.55$), disturbed clusters exhibit a more pronounced faint end slope than their relaxed counterparts. It was hypothesized that galaxy cluster populations show differences between relaxed and disturbed clusters at $z \gtrsim 0.55$, which was corroborated by \citet{Aldas+23}, at least in the color-magnitude relation of these systems.

In this series of two papers, we aim to investigate the impact of the dynamical state of massive clusters ($M_{500} \geq 1.5 \times 10^{14}$ M$_{\odot}$) on the physical and morphological properties of their member galaxies, in the redshift range $0.10 < z < 0.35$. In this first paper, our aim is to detail the sample selection and use optical data from the DESI Legacy Imaging Survey Data Release 10 and X-ray data from the Chandra and XMM-Newton archives to determine the dynamic state of the clusters. The calculation of the physical and morphological properties of the galaxies, as well as the analysis of the impact of the dynamical state on these galaxy populations, will be presented in a second article (Véliz Astudillo et al., in prep, hereafter Paper II).

This paper is organized as follows. Section \ref{sec:data} describes the data and the sample selection process. In Section \ref{sec:cluster-membership}, we detail the method used to assign cluster memberships, followed by the approach used to select the BCGs in Section \ref{sec:bcg-selection}. We define the dynamical state proxies in Section \ref{sec:dyn-state-proxies}. In Section \ref{sec:dyn-states}, we classify the dynamical state of galaxy clusters, explaining the chosen method. We discuss the results and present the conclusions in Section \ref{sec:discussion}.

Throughout this work, we adopt a flat $\Lambda$CDM cosmology, assuming $H_0 = 69.3$ km s$^{-1}$, $\Omega_{\Lambda} = 0.721$ and $\Omega_{m} = 0.287$ \citep[WMAP-9,][]{Hinshaw+13}. Unless specifically stated, the magnitudes used in this article are quoted in the AB system.

\section{Data}
\label{sec:data}

\subsection{DESI Legacy Survey Data Release 10}

The DESI Legacy Imaging Survey (Legacy Survey, LS) is the result of the efforts of three different surveys that provide images and catalogs, initially planned for the $g$, $r$, and $z$ filters. On the one hand, there are surveys in the northern hemisphere ($\delta \gtrsim 32$); these include the Beijing Arizona Sky Survey \citep[BASS,][]{Zou+17}, which provides photometric information in the $g$ and $r$ bands using the Bok 2.3 m Telescope, and the Mayall $z$-band Legacy Survey \citep[MzLS,][]{Silva+16}. On the other hand, for the southern hemisphere, there is the Dark Energy Camera Legacy Survey \citep[DECaLS,][]{Blum+16}, which makes use of the Dark Energy Camera (DECam) mounted on the Victor Blanco 4 m Telescope in the $g$, $r$, and $z$ filters. It is worth mentioning that DECaLS is primarily composed of data from the Dark Energy Survey \citep[DES,][]{DESCollab+05}, with the remaining data obtained from the public DECam data, available in the NOIRLab online repository. Among these earlier programs, the DECam Local Volume Exploration Survey \citep[DELVE,][]{DrlicaWagner+22}, and the DECam eROSITA Survey (DeROSITAS; PI: Zenteno). The initial goal of LS was to select targets for the upcoming DESI Spectroscopic Survey, whose spectrograph will be installed on the 4m Mayall Telescope, precisely where the MzLS is conducted.

The latest data release (LS DR10) included DECam $i$ band data. Combined, all images from LS DR10 cover more than 20,000 square degrees of the sky, with the three surveys having similar observational design properties. In each filter, a uniform depth of $5\sigma$ is reached at $m_g = 24.7$, $m_r = 23.9$, $m_z = 23.0$, which is typically $\sim 1$ magnitude deeper than Pan-STARRS \citep{Duncan22}. Detailed descriptions of source detection and photometry can be found in \citet[][and references therein]{Dey+19}.

On the official LS DR10 website\footnote{\url{https://www.legacysurvey.org/dr10/}}, catalogs and images are available. Regarding the catalogs, they were generated using the software \texttt{The Tractor} \citep{Lang+16}. This code employs a probabilistic method to fit surface brightness models to the sources in an image. The models can be point sources (PSF), round exponential galaxies with variable radius (REX), de Vaucouleurs (DEV) profiles (elliptical galaxies), exponential (EXP) profiles (spiral galaxies), and Sersic (SER) profiles, and these are fitted employing a $\chi^2$ minimization problem. It is worth noting that in addition to optical data, the catalogs also include infrared information from the Wide-field Infrared Survey Explorer (WISE) five-year coadded image, also known as unWISE \citep{Wright+10}, in the $W1$, $W2$, $W3$, and $W4$ bands. Concerning the images, we utilized science images, weight images, exposure number maps, and full-width half-maximum maps.

\subsection{Chandra and XMM-Newton archive images}

The X-ray images used in this study were obtained from the Chandra and XMM-Newton archives. However, it is important to note that the entire process, from the downloading of the images to their uniform processing and availability in repositories, was carried out by other authors and is accessible to the scientific community. Specifically, \citet{Yuan&Han20} initiated their project to evaluate the dynamical state of galaxy clusters based on the morphology of the X-ray surface brightness distribution, revealing the behaviors of the ICM in these systems. In their study, they explain that the Chandra satellite has observed approximately $\sim 1000$ galaxy clusters, and the high resolution of its images is ideal for detecting substructures in the ICM. They processed the data from these structures by filtering photons with energies in the 0.5$-$5.0 keV range, carefully removing flares and point sources, and smoothing the images with a Gaussian kernel at a physical scale of 30 kpc. Initially, these scales varied for each cluster due to their different redshifts.

Subsequently, \citet{YuanHan&Wen22} expanded their previous work, this time including data from $\sim 1300$ XMM-Newton systems and 22 new Chandra-observed clusters in the period between their two studies. The processing of X-ray images was the same for consistency, resulting in a total of 1844 galaxy clusters between the two studies. In addition, parameters were calculated for all these clusters to approximate their dynamical states, which will be detailed in Section 3.3. Thus, the final data products, consisting of images and catalogs, are publicly available on the authors' web repository\footnote{\url{http://zmtt.bao.ac.cn/galaxy_clusters/dyXimages/}}.

\subsection{Redshift catalogs}

This study uses the photometric redshift catalog from \citet{Wen&Han22}. Using data from DES and unWISE, they estimated the photometric redshifts of 105 million galaxies, employing clustering algorithms to identify 151,244 galaxy clusters in the redshift range of $0.1< z < 1.5$. The algorithms used are explained in detail in \citep{Wen&Han21}. In summary, the authors used publicly available DES DR2 data in the $grizY$ bands \citep{Abbott+21} and unWISE data in the $W1$ and $W2$ filters. The match between both databases results in 105 million objects. Then, photometric redshifts were derived on the basis of colors using the nearest-neighbors algorithm with respect to the spectroscopic redshifts of a robust training sample. Specifically, the distances from the target galaxies to all training galaxies in the multidimensional color space ($g-r$, $r-i$, $i-z$, $z-y$, $y-W1$, and $W1-W2$) were calculated. Then, the 20 nearest neighbors were selected to compute the photometric redshift. This was calculated as the mean value of these 20 neighbors with spectroscopic redshift and the error was calculated as the standard deviation among them. This methodology is based on the premise that the color of galaxies is closely related to their redshifts when covering spectral features such as the 4000\AA~ break or the Balmer jump. Galaxies close to the multidimensional color space generally have similar redshifts \citep{Wen&Han21}.

We conducted an exhaustive literature search to find galaxy clusters with spectroscopic information in their respective fields. The selection process of these systems is detailed in Section \ref{sec:sample-selection}, and Table \ref{tab:spectroscopic-catalogs} shows the sources from which the spectroscopic information was obtained.

\begin{table}
    \caption{Galaxy clusters with spectroscopic redshifts available in the literature.}
    \label{tab:spectroscopic-catalogs}
    \centering
    \resizebox{0.45\textwidth}{!}{
    \begin{tabular}{lr}
    \hline\hline
    Cluster & References\\
    \hline
    Abell 222 & \citet{Proust+00} \\
    Abell 223 & \citet{Proust+00} \\
    Abell 267 & \citet{Rines+13, Tucker+17}\\
    Abell 383 & \citet{Geller+14}\\
    Abell 402 & \citet{Richard+21} \\
    Abell 2537 & \citet{Braglia+09}\\
    Abell 2631 & \citet{Rines+13}\\
    Abell 2744 & \citet{Braglia+09,Owers+11,Richard+21}\\
    Abell 2813 & \citet{Guzzo+09} \\
    Abell 3088 & \citet{Guzzo+09} \\
    Abell 3364 & \citet{Guzzo+09} \\
    Abell 3378 & \citet{Guzzo+09} \\
    Abell 3739 & \citet{Guzzo+09} \\
    Abell 3827 & Carrasco et al. (in prep)\\
    Abell S295 & \citet{Ruel+14,Bayliss+16}\\
    Abell S520 & \citet{Guzzo+09,Foex+17} \\
    Abell S579 & \citet{Guzzo+09} \\
    Abell S1063 & \citet{Mercurio+21}\\
    ACT-CLJ0235-5121 & \citet{Sifon+16} \\
    RBS 1748 & \citet{Sifon+16}\\
    RXC J0117.8-5455 & \citet{Guzzo+09} \\
    RXC J0220.9-3829 & \citet{Guzzo+09} \\
    RXC J0528.2-2942 & \citet{Guzzo+09} \\
    RXC J0532.9-3701 & \citet{Guzzo+09} \\
    RXC J2011.3-5725 & \citet{Guzzo+09} \\
    RXC J2023.4-5535 & \citet{Guzzo+09} \\
    SPT-CLJ0106-5943 & \citet{Bayliss+16}\\
    SPT-CLJ0348-4514 & \citet{Bayliss+16} \\
    SPT-CLJ2032-5627 & \citet{Ruel+14}\\
    SPT-CLJ2130-6458 & \citet{Ruel+14} \\
    SPT-CLJ2138-6007 & \citet{Ruel+14}\\
    ZwCl 2341.1+0000 & \citet{Boschin+13} \\
    \hline
    \end{tabular}
    }
\end{table}

\begin{figure*}[h]
    \centering
    \includegraphics[width=1\textwidth]{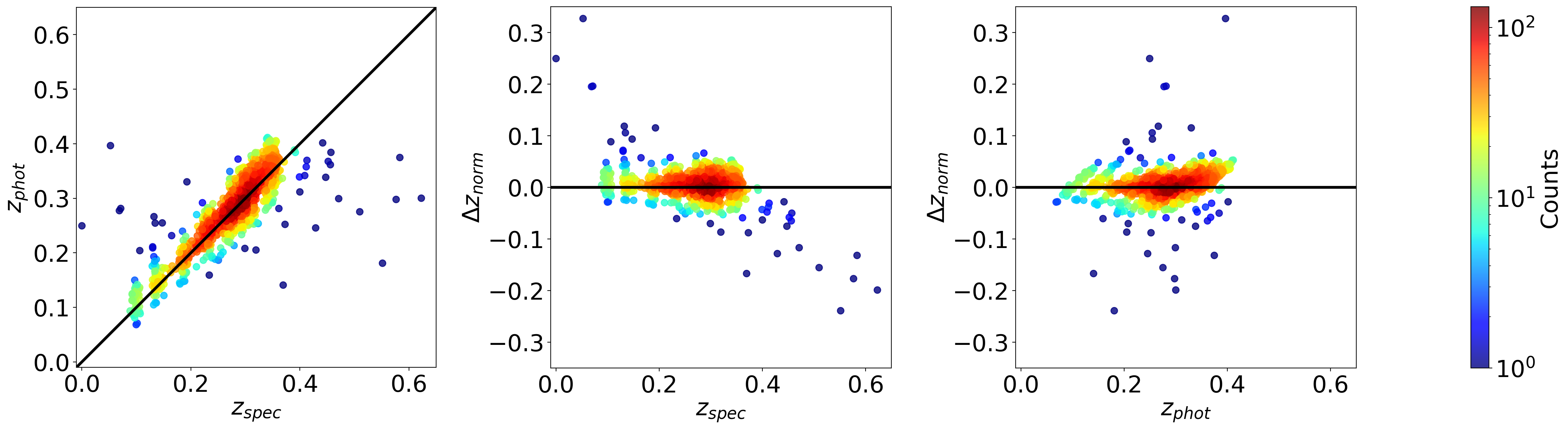}
    \caption{Left: Comparison between $z_\text{spec}$ and $z_{\text{phot}}$. The solid black line represents the identity line ($z_{\text{spec}} = z_{\text{phot}}$). Middle: $\Delta z_{\text{norm}}$ as function of $z_{\text{spec}}$. Right: $\Delta z_{\text{norm}}$ as function of $z_{\text{spec}}$.  In the latter two cases, the solid black line corresponds to the constant function where $\Delta z_{\text{norm}} = 0$.}
    \label{fig:zphot-zspec-comparison}
\end{figure*}

Fig. \ref{fig:zphot-zspec-comparison} shows the comparison between the photometric redshifts obtained from the catalogs of \citet{Wen&Han22} and the spectroscopic redshifts obtained in the literature for the member galaxies of the clusters. We define $\Delta z_{\text{norm}} = (z_{\text{phot}} - z_{\text{spec}}) / (1 + z_{\text{spec}})$. We found a mean value of $\Delta z_{\text{norm}} = 0.005$ with a dispersion of $\sigma_{\text{z}} = 0.02$, using robust biweight estimators.

\subsection{Sample selection}
\label{sec:sample-selection}

If we want to study the effect of cluster dynamics on the morphology of their member galaxies, having a robust and homogeneous sample is crucial. To achieve this, we performed a 2-arcminute radius cross-match between the galaxy clusters identified in the optical by \citet{Wen&Han22} and those in X-rays found in the Chandra \citep{Yuan&Han20} and XMM-Newton \citep{YuanHan&Wen22} archives, resulting in 471 systems.

Then, we selected those galaxy clusters in the redshift range of $0.10 < z < 0.35$, resulting in 152 clusters. At this point, Abell 3827 is added, which is at $z \sim 0.099$, and we have access to spectroscopic information. The lower limit is chosen mainly because of the extensive spatial coverage required for nearby clusters. Regarding the upper limit, we rely on the work of \citet{DeAlbernazFerreira&Ferrari18}, demonstrating that reliable morphological studies can be performed using DECam up to $z \sim 0.4$. However, we decided to be conservative and study up to the mentioned limit. Furthermore, within this interval, it ensures that we cover the 4000 \AA\ break using the $(g-r)$ color index to separate early-type red galaxies from star-forming objects \citep[e.g.,][]{Bruzual83,NiloCastellon+14}. Moreover, within this redshift range, the LS DR10 achieves enough photometric depth to conduct homogeneous studies up to three magnitudes fainter than the characteristic magnitude ($m^* + 3$) of all clusters in the sample.

Subsequently, to ensure that the matching of clusters between optical and X-ray databases is not affected by projection effects, we impose the condition that $z_x - z_{\text{phot}} / (1 + z_x) \leq 0.05$, where $z_x$ is the redshift found in the X-ray catalogs of \citet{Yuan&Han20} and \citet{YuanHan&Wen22}, while $z_{\text{phot}}$ is the photometric redshift estimated for the clusters identified in the optical wavelengths by \citet{Wen&Han22}, resulting in 119 galaxy clusters. Next, we remove from the sample all systems that are very poor and low-mass by applying the condition $\lambda \geq 30$ and $M_{500} \geq 1.5 \times 10^{14} M_{\odot}$, where $\lambda$ is the cluster richness estimated based on overdensities and fluctuations in the field, and $M_{500}$ is the mass of the clusters estimated within $R_{500}$, with calculations detailed in \citet{Wen&Han22}. Finally, the last filter was to exclude from the sample, via visual inspection, all systems with field contamination due to a saturated star or lack of photometric information in the LS catalogs. Our final sample consists of 87 massive galaxy clusters.

\section{Cluster membership}
\label{sec:cluster-membership}

The most unequivocal way to select member galaxies of a cluster is by using spectroscopic redshifts. However, for our sample, less than half of the clusters have catalogs of this nature available in the literature or databases such as NED\footnote{\url{https://ned.ipac.caltech.edu/}} or Vizier\footnote{\url{https://vizier.cds.unistra.fr/}}. Additionally, there is the difficulty of the few systems with this information not sharing the same observational designs. For example, some systems have more than 1,000 galaxies with spectroscopic data in the field due to the use of integral field spectroscopy \citep[e.g.,][]{Mercurio+21} or multiobject spectroscopy \citep[e.g.,][]{Owers+11}. In contrast, others have fewer than 20 objects due to the use of a few slits, mainly selecting galaxies that are on the Red Sequences (RS) in the color-magnitude diagrams (CMD) of the clusters \citep[e.g.,][]{Bayliss+16}. Additionally, there are differences in sky coverage and/or limit magnitudes. These differences can directly affect sample analyses, leading to biased results due to inhomogeneities. However, this can be addressed by homogeneously assigning membership to the entire sample using only photometric redshifts.

In this way, we follow a probabilistic method whose foundations were established by \citet{Brunner&Lubin00} and later developed by \citet{Pello+09}. The procedure consists of calculating the probability $P_{\text{member}}$ that a galaxy is a member of a cluster at redshift $z_{\text{cl}}$ within a redshift slice $\delta_z$.

\begin{equation}
    P_{\text{member}} = \int_{z_{\text{cl}} - \delta_z}^{z_{\text{cl}} + \delta_z} P(z) dz.
\end{equation}

To define the redshift slice, we follow \citet{Kesebonye+23}, where they use $\delta_z = n \sigma_{\text{bw}} (1 + z_{\text{cl}})$, with $n=2$. To determine $\sigma_{\text{bw}}$, we used all galaxies with available spectroscopic redshifts (see Table \ref{tab:spectroscopic-catalogs}) to compare them with photometric redshifts, obtaining a mean residual of $(z_{\text{spec}} - z_{\text{phot}}) / (1+z_{\text{spec}}) = 0.008$. The robust biweight scale estimator \citep{Beers+90} resulted in $\sigma_{\text{bw}} = 0.02$.

$P(z)$ is the galaxy photometric redshift probability distribution. This distribution has been discussed in the literature, which does not have to be Gaussian; instead, it can have multiple peaks or extended tails \citep{Pello+09}. For this reason, the photometric redshift distributions provided as output in codes that calculate this value have been used in other works \citep[e.g.,][]{DeLucia+04,Vulcani+12,Kesebonye+23}. These distributions are usually obtained using algorithms that involve fitting SEDs to a library of templates. However, we do not have this information; we only have the photometric redshift information and its corresponding error. Thus, we approximate this distribution as a Gaussian with $\mu = z_{\text{phot}}$ and $\sigma = z_{\text{phot},\text{err}}$. Although we recognize the mentioned disadvantages, at least we consider the measurement error, and it is feasible to apply this probabilistic method instead of making a simple selection within a photometric redshift slice.

The method requires a calibration of $P_{\text{member}}$ with spectroscopic data, which, in turn, allows the calculation of the completeness of cluster member galaxies and contamination by field galaxies. We perform this calibration using clusters that have spectroscopic redshifts for a large number of both member and field galaxies, specifically Abell 267, Abell 383, Abell 2537, Abell 2631, Abell 2744, Abell S1063, and RBS 1748. We select spectroscopically confirmed member galaxies using $|z_{\text{cl}} - z_{\text{spec}}| < 3 \sigma_{\text{cl}} (1+z_{\text{cl}})$, where $\sigma_{\text{cl}}$ is the velocity dispersion of the cluster.

\begin{figure}[h]
    \centering
    \includegraphics[width=0.45\textwidth]{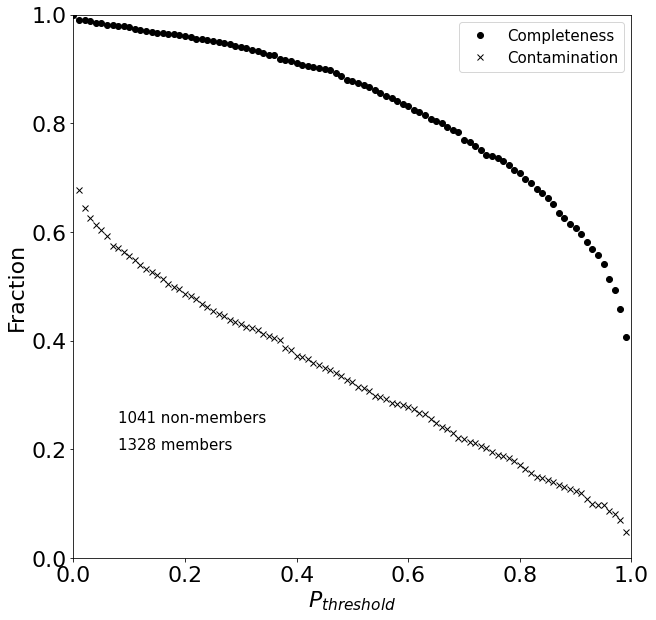}
    \caption{Fraction of retained members (completeness, black dots) and non-members (contamination, gray crosses) as a function of the probability threshold $P_{\text{threshold}}$. This Fig. is created using only galaxies with secure $z_{\text{spec}}$ measurements in clusters with enough spectroscopic data for members and the field.}
    \label{fig:membership-calibration}
\end{figure}

We identify 1,041 member galaxies and 1,328 galaxies that do not belong to the mentioned clusters. In Fig. \ref{fig:membership-calibration}, the completeness and contamination fractions can be observed for a range of $P_{\text{threshold}}$ values. We explored 100 variations for this parameter, from 0 to 1, in steps of 0.01. With this calibration, we decided to use the value of $P_{\text{threshold}} = 0.65$, ensuring a completeness of at least 80\% and contamination below 25\%\footnote{Completeness and contamination are calculated by selecting a membership probability threshold based on photometric data, and these values are obtained by comparison with spectroscopic data and assumed for the entire sample.}. This photometric redshift quality, completeness, and contamination levels work well for our purposes, as we plan to study the integrated properties of cluster galaxy populations, rather than the individual properties of galaxies. Moreover, as we explain in Section \ref{sec:dyn-state-proxies}, we use X-ray data and the position of the BCG derived from optical images to evaluate the dynamical state of galaxy clusters, and we do not aim to find substructures using member galaxies selected with photometric redshifts.

Also, we apply a magnitude limit to the study sample. We use a criterion of $m < m^* + 3$ in the $r$ band. This ensures a homogeneous selection of members for all clusters in the whole redshift range.

\begin{figure}
    \centering
    \includegraphics[width=1.0\linewidth]{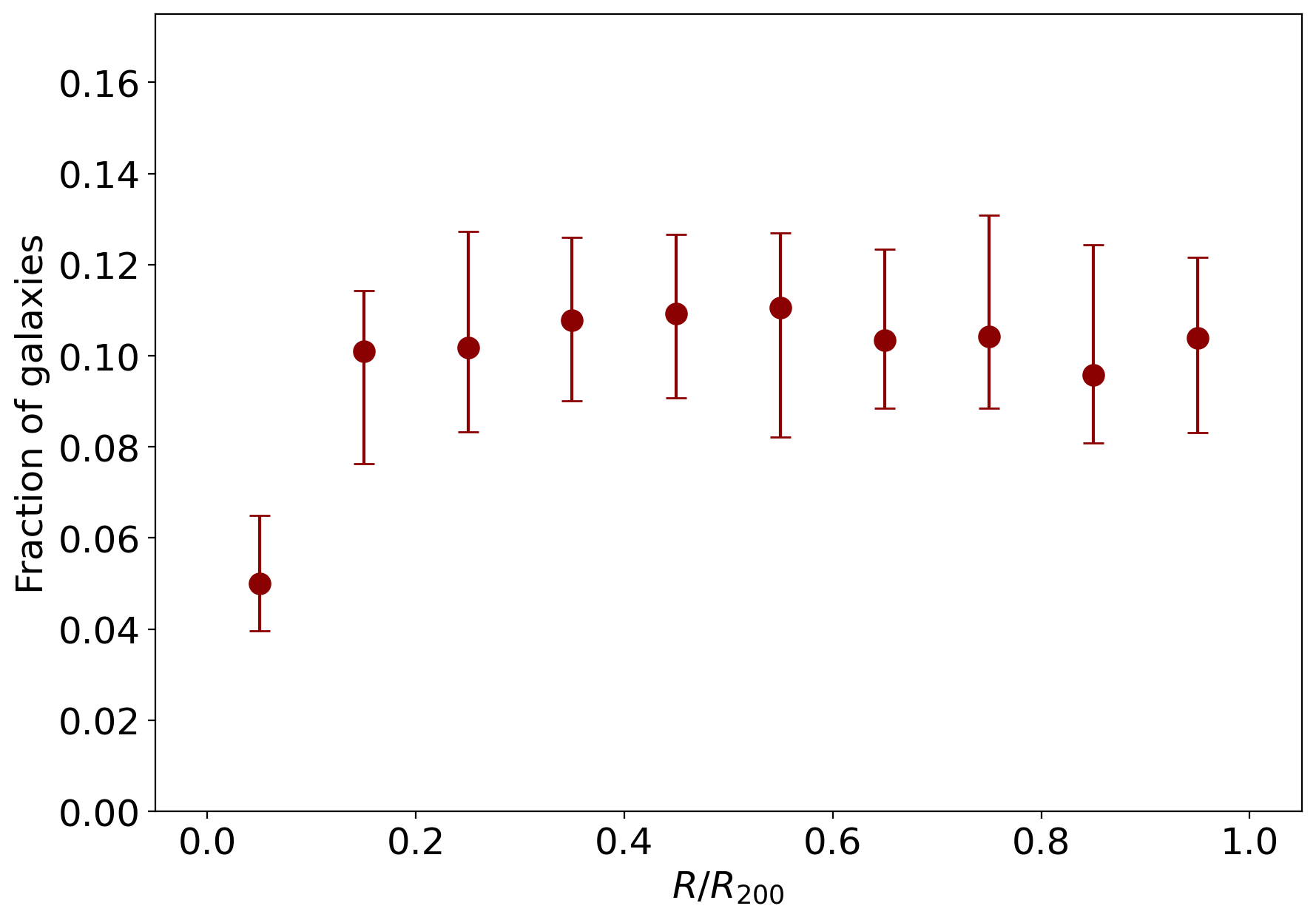}
    \caption{Fraction of galaxies as function of the clustercentric distance for the stacked sample of galaxy clusters. Dark red points correspond to the median values in each bin, while the lower and upper error bars are the first and third quartile in each bin, respectively.}
    \label{fig:spatial-coverage-clusters}
\end{figure}

In Fig. \ref{fig:spatial-coverage-clusters} we show the spatial coverage of the selected members in $R_{200}$. We split the clustercentric distance in 10 symmetric bins, and we present the median values along with the first and third quartiles, as a measure of dispersion. It can be seen that the variations are very small (the maximum dispersion is lower than $3\%$). This means that the clusters do not present any bias in the spatial distribution and, therefore, are indeed comparable.

\section{BCG selection}
\label{sec:bcg-selection}

Identifying the BCG is crucial for estimating the dynamical state of galaxy clusters. We select the BCG using a combination of an automatic method and visual inspection. 

The automatic method involves selecting the brightest cluster member galaxy within $\pm 1\sigma$ of the best fit to the red cluster sequence in the normalized color-magnitude diagram of each galaxy cluster. Several studies have shown that the evolution of the characteristic magnitude $m^*$ for galaxy cluster populations can be described by a stellar population formed at high redshift that evolves passively \citep[e.g.,][]{DePropris+99,Andreon06,Stalder+13}. This has been reproduced over a broad redshift range ($0 \lesssim z \lesssim 1$) for massive clusters \citep{Zenteno+16,Zenteno+20}. To model $m^*$, we created red sequence composite stellar populations (CSP) for the DECam \textit{griz} bands using \citet{BruzualCharlot03} SSP models via the Python code \textsc{EzGal} \citep{Mancone&Gonzalez12}. Specifically, we assumed an exponentially declining star formation rate with a characteristic decay time of 0.4 Gyr, the Chabrier initial mass function (IMF), and a single burst of star formation at $z=3$ followed by passive evolution to $z=0$. Next, we define normalized magnitudes as

\begin{equation}
    m_{\text{norm}} = m - m^*,
\end{equation}

where $m$ is the magnitude of each galaxy. This allows us to compare the CMDs of galaxy clusters despite differences in redshift as we consider the cluster redshift and K-corrections\footnote{The K-corrections are applied to the CSP that models the evolution of $m^*$. As a result, the same $m^*$, specific to each cluster and already incorporating the K-correction, is used for all galaxies within that cluster, regardless of their morphology.} in the applied models. Note that we inherently define normalized colors that correspond to the subtraction of two normalized magnitudes.

\begin{figure*}[h]
    \centering
    \includegraphics[width=1\textwidth]{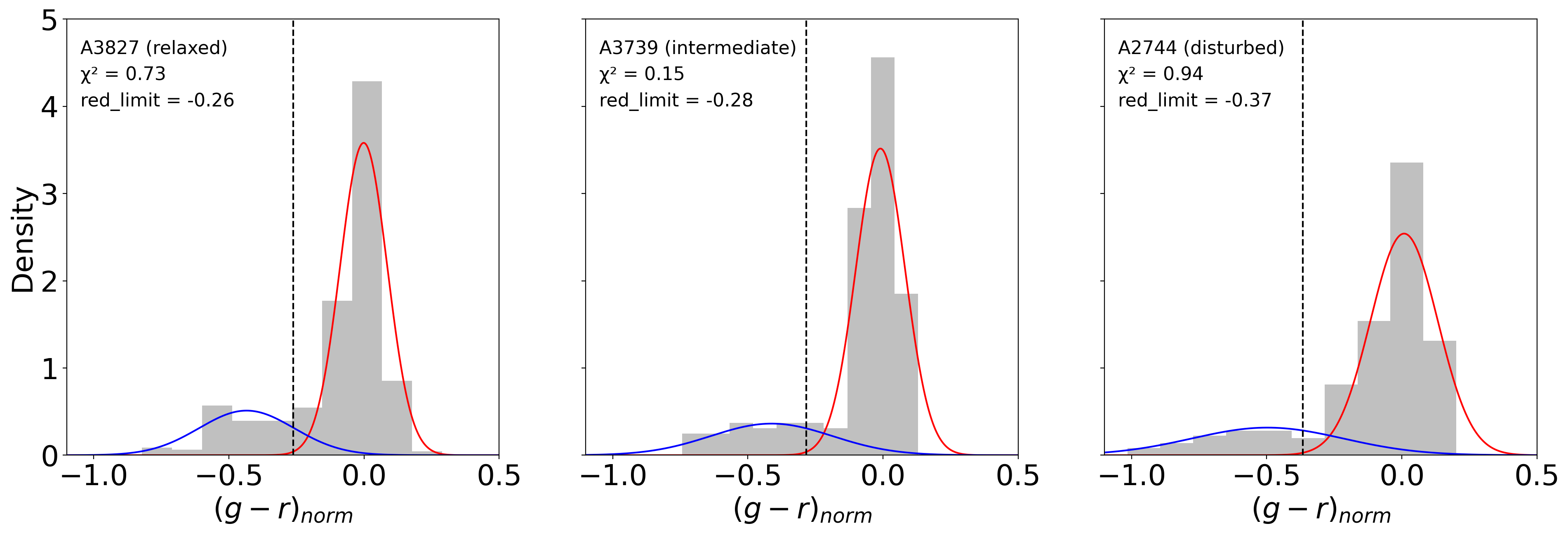}
    \caption{Double Gaussian fits applied to the normalized color distribution of each cluster. From left to right panels, Abell 3827 (relaxed cluster), Abell 3739 (intermediate cluster), and Abell 2744 (disturbed cluster) are presented as examples. The colors of the Gaussians are associated with the red and blue components of the galaxy clusters. The black dashed line indicated the \textit{red\_limit} in each panel. The specific value of this parameter is located in the upper left corner of each panel along with the $\chi^2$ statistic for each fit performed with \texttt{lmfit}.}
    \label{fig:double-gaussian-fits}
\end{figure*}

Subsequently, we fit a double Gaussian model to the normalized color distribution $(g-r)_{\text{norm}}$ of each galaxy cluster using a magnitude limit of $r_{\text{norm}} = 3$, i.e., $r = m^* + 3$. This is based on the well-known bimodal color distribution in the member galaxies of these systems, featuring both the blue cloud and the red sequence \citep[e.g.,][]{Baldry+04,Balogh+04,Menci+05}. We define the \textit{red\_limit} as $\mu_{\text{red}} - 3 \sigma_{\text{red}}$, where $\mu_{\text{red}}$ and $\sigma_{\text{red}}$ correspond to the mean value and standard deviation of the Gaussian fitted to the red component, respectively. An example of these double Gaussian fittings for a cluster for each dynamical state is shown in Fig. \ref{fig:double-gaussian-fits}.

\begin{figure*}[h]
    \centering
    \includegraphics[width=1\textwidth]{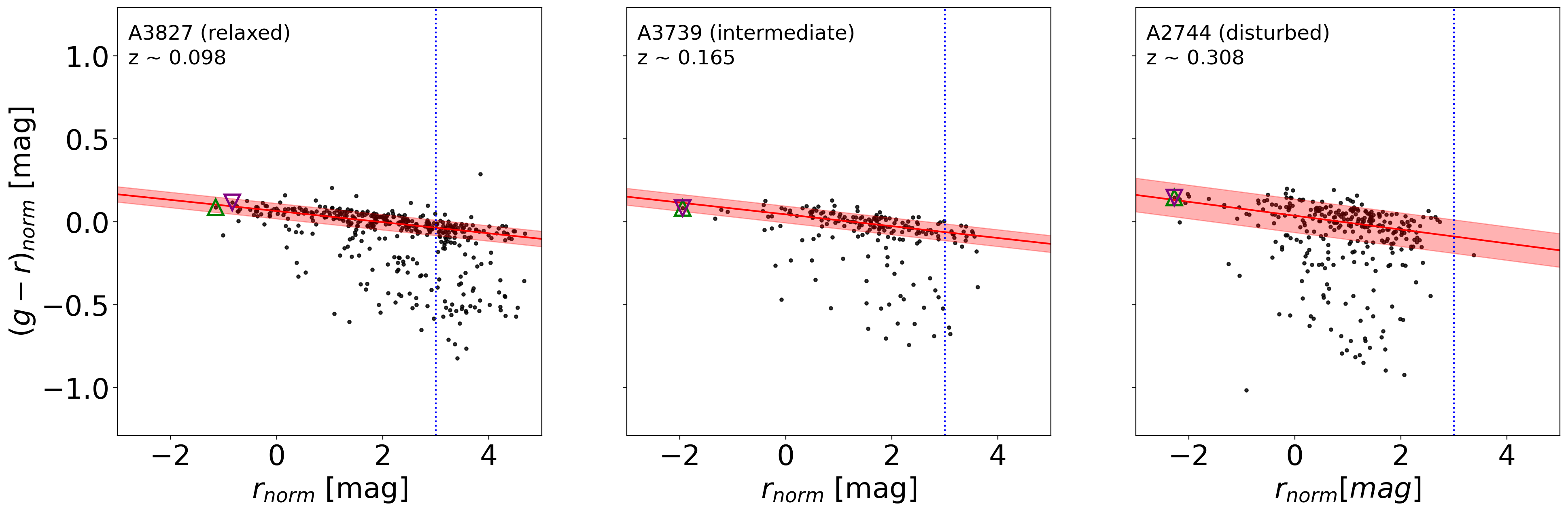}
    \caption{Normalized color-magnitude diagrams for the same clusters mentioned in Fig. \ref{fig:double-gaussian-fits}. Black dots correspond to galaxies selected as members using the probabilistic method with photometric redshifts. Green triangles represent galaxies pre-selected as BCGs using the automated method, while purple inverted triangles denote visually confirmed BCGs. The solid red lines represent the best fits of the red sequences calculated with HuberRegressor, and the red shaded areas are the 1$\sigma$ regions of them. The vertical blue dashed lines indicate the magnitude limit for studying physical and structural properties, i.e., $m^*+3$, where $m^*$ is the characteristic magnitude at each cluster redshift modeled with CPS.}
    \label{fig:red-sequence-fits}
\end{figure*}

After that, we select the red galaxies from each galaxy cluster as those where $(g-r)_{\text{norm}} >$ \textit{red\_limit} and apply a robust linear regression using the HuberRegressor \citep{Huber11} to fit the Red Sequence (Fig. \ref{fig:red-sequence-fits}). This regression model has the advantage of being less influenced by the presence of outliers. To achieve this, the sample is divided into two groups, with inliers having an absolute error smaller than a certain threshold. Those that do not meet this condition are considered outliers and given less weight. Specifically, the HuberRegressor optimizes the squared loss for the sub-sample where $|(y - X_w - c) / \sigma| < \epsilon$ (inliers) and the absolute loss for the sub-sample where $|(y - X_w - c) / \sigma| > \epsilon$ (outliers), where the model coefficient $w$, intercept $c$ and scale $\sigma$ are the parameters to optimize. To achieve a statistical efficiency of 95\%, this threshold $\epsilon$ was set to 1.35. The brightest galaxies within $1\sigma$ of the best RS fit of each cluster are selected as BCG candidates.

Finally, we visually inspect the BCGs to confirm that the selection was correct. To do this, we follow the approach of \citet{Zenteno+20}, considering several properties, including size, colors, the number of neighboring galaxies, and their proximity to X-ray peaks.

It is important to note that the method that we ultimately use to discern the BCG of each system is visual inspection, which is particularly useful for measuring the accuracy of the automatic process. However, it is crucial to ensure with spectroscopic information that the galaxy selected as the BCG truly belongs to the cluster. For more details on the accuracy of both methods, refer to Section \ref{sec:disc-bcg-selection-accuracy}.

\section{Proxies of dynamical states}
\label{sec:dyn-state-proxies}

In this subsection, we define the six dynamical state proxies that we utilize. The first two correspond to the offset between the BCG and the X-ray peak and centroid, and are calculated directly by us. The next four were calculated by \citet{Yuan&Han20} and \citet{YuanHan&Wen22}, and we extracted the values from their catalogs. The choice to use six parameters together is because each has its limitations, as we describe below, and their combination can provide us with more robust results.

\subsection{BCG-X-ray centroid/peak offset}

Considering the highly hydrodynamically collisional nature of the hot gas composing the ICM of galaxy clusters and the typical positioning of the BCG within the deepest region of its gravitational potential well, the BCG can be effectively used as a proxy for the non-collisional dark matter component. Thus, the offset between the position of the X-ray emission peak/centroid and the BCG provides a good approximation of the dynamical state of clusters, at least in the projected sky plane. This is because the components would be noticeably displaced in a merger process between these structures.

Following \citet{Mann&Ebeling12}, we use a threshold to distinguish between relaxed and disturbed clusters, with a value of $D_{\text{BCG-X}} = 42(71)$ kpc for the X-ray peak (centroid) offset.

\subsection{Morphological parameter $\delta$}

This parameter is a morphological indicator of the X-ray surface brightness distribution. The definition of \citet{Yuan&Han20} mentions that the great advantage of this dynamical parameter is its adaptability to the properties of each cluster, allowing a direct comparison between them. The first step in calculating the value of $\delta$ is to fit a 2D elliptical $\beta$ model to the X-ray surface brightness distribution. Then, a new parameter of the X-ray distribution profile is calculated:

\begin{equation}
    \kappa = \frac{1 + \epsilon}{\beta},
\end{equation}

where $\beta$ and $\epsilon$ are the power-law index and ellipticity of the fitted $\beta$-model, respectively.

Based on the observation that disturbed clusters tend to have a more asymmetric geometry than relaxed clusters \citep[e.g.,][]{Okabe+10, Zhang+10}, the asymmetry factor $\alpha$ is used as an auxiliary variable.

\begin{equation}
    \alpha = \frac{\sum_{x_i, y_i} [f_{\text{obs}} (x_i, y_i) - f_{\text{obs}} (x_i', y_i')]^2}{\sum_{x_i, y_i} f^2_{\text{obs}} (x_i, y_i)} \times 100 \text{ per cent},
\end{equation}

where, $f_{\text{obs}} (x_i', y_i')$ is the flux observed in the pixel symmetric to $f_{\text{obs}} (x_i, y_i)$ with respect to the center of the cluster $(x_0, y_0)$, which is obtained from the fit of the $\beta$-model.

With a combination of these parameters that quantify the properties of the profiles and the asymmetry of the X-ray surface brightness distribution, the morphological index $\delta$ is defined as:

\begin{equation}
    \delta = A \kappa + B \alpha + C.
\end{equation}

To find appropriate values for the free parameters, \citet{Yuan&Han20} used a homogeneous calibration sample of 125 clusters qualitatively classified by \citet{Mann&Ebeling12}, finding that the values of $A=0.68$, $B=0.73$, and $C=0.21$ can separate the sample into relaxed systems ($\delta < 0$) and disturbed systems ($\delta \geq 0$) with a success rate of 88\%.

\subsection{Concentration $c$}

Galaxy clusters in a non-virialized state may exhibit various substructures or extended geometries. However, when these systems are very close to virialization, most of the matter is concentrated in their center, with some even having very luminous cool cores \citep[e.g.,][]{Fabian94,McDonald+12,McDonald+13}. The concentration index, $c$, quantifies this characteristic by being calculated as the ratio of the integrated X-ray flux within an inner aperture to that within an outer aperture. The definition of \citet{Cassano+10, Cassano+13} was followed, which used apertures of 100 kpc and 500 kpc:

\begin{equation}
    c = \frac{S_{100\text{ kpc}}}{S_{100\text{ kpc}}} = \frac{\sum_{R < 100\text{ kpc}} f_{\text{obs}}(x_i,y_i)}{\sum_{R < 500\text{ kpc}} f_{\text{obs}}(x_i, y_i)},
\end{equation}

where $f_{\text{obs}} (x_i, y_i)$ is the flux observed in the pixel $(x_i, y_i)$.

\citet{Cassano+10} found that the median distribution of the concentration parameter $\log(c) = -0.7$ is a reasonable threshold to distinguish galaxy clusters in different dynamical states. These states were classified based on their features observed in radio images.

\subsection{Centroid shift $\omega$}

In relaxed clusters, there is an almost negligible deviation between the positions of the X-ray peak and the X-ray centroid. However, these two positions can exhibit a considerable shift in clusters that undergo mergers. \citet{Poole+06} quantified the deviation between the X-ray peak and the center of a model fit within 20 apertures of different sizes centered on the X-ray peak. The aim was to obtain a value that helps determine the dynamical state of the cluster. This is an iterative process that starts from 0.05 $R_{\text{ap}}$ and increases to $R_{\text{ap}}$ in increments of $0.05R_{\text{ap}}$. In other words,

\begin{equation}
    \omega = \left[ \frac{1}{n-1} \sum_i (\Delta_i - \langle \Delta \rangle)^2 \right]^{\frac{1}{2}} \times \frac{1}{R_{\text{ap}}}.
\end{equation}

Here, $R_{\text{ap}} = 500$ kpc, $n = 20$, $\Delta_i$ is the distance between the peak of the X-ray surface brightness and the center of the fitted model in the i-th aperture, and $\langle \Delta \rangle$ is the mean value of all $\Delta_i$.

Similarly to concentration $c$, \citet{Cassano+10} found that $\log(\omega) = -1.92$ is a good threshold for separating relaxed and disturbed systems.

\subsection{Power ratio $P_3/P_0$}

Based on the observation that systems undergoing mergers or with substructures exhibit more fluctuations in X-ray surface brightness, \citet{Buote&Tsai95} derived dimensionless powers from the 2D multipole expansion of the projected gravitational potential of clusters within $R_{\text{ap}} = 500$ kpc.

\begin{equation}
    P_0 = [a_0 \ln(R_{\text{ap}})]^2,
\end{equation}

for $m = 0$, and

\begin{equation}
    P_m = \frac{1}{2m^2R_{\text{ap}}^{2m}} (a^2_m+b^2_m),
\end{equation}

for $m > 0$.

Here, the moments $a_m$ and $b_m$ are given by

\begin{equation}
    a_m = \int_{r \leq R_{\text{ap}}} f_{\text{obs}}(x_i, y_i)(r)^m \, \cos(m\phi) \, dx_i \, dy_i,
\end{equation}

\begin{equation}
    b_m = \int_{r \leq R_{\text{ap}}} f_{\text{obs}}(x_i, y_i)(r)^m \, \sin(m\phi) \, dx_i \, dy_i,
\end{equation}

where $r$ is the distance to the pixel $(x_i, y_i)$ from the center of the cluster ($x_0, y_0$), and $f_{\text{obs}} (x_i, y_i)$ is the flux observed in that pixel.

It has been empirically found that the power ratio $P_3/P_0$ is a good parameter to separate clusters according to their dynamical states. In this work, we use $\log(P_3/P_0) = -6.92$ as a threshold, considering clusters with a value lower (higher) than this threshold as relaxed (disturbed) systems. This choice, like the parameters $c$ and $\omega$, is based on the article by \citet{Cassano+10}, where the thresholds are defined as the medians of the distributions.

\section{Dynamical states of galaxy clusters}
\label{sec:dyn-states}

As mentioned earlier, dynamical parameters alone do not robustly separate galaxy cluster samples. This is due to various factors appropriately discussed in the articles in which they were developed \citep[i.e.,][]{Buote&Tsai95,Poole+06,Cassano+10,Mann&Ebeling12,Yuan&Han20}. For example, parameters such as $c$, $\omega$, and $P_3/P_0$ are calculated on fixed apertures. However, clusters have different sizes, so the values derived from these apertures may not be entirely comparable. That is why \citet{Yuan&Han20} decided to develop a new parameter ($\delta$), which is adaptable to the size of these systems. Despite this, the parameter alone still has a considerable overlapping region.

If we add the offsets of the BCG with the X-ray peak and centroid to the parameters as mentioned earlier, which are pretty robust proxies due to their multiwavelength nature and the components they approximate in each structure, then we propose that we can more robustly approximate the dynamical state of galaxy clusters.

\begin{figure*}[h]
    \centering
    \includegraphics[width=1\textwidth]{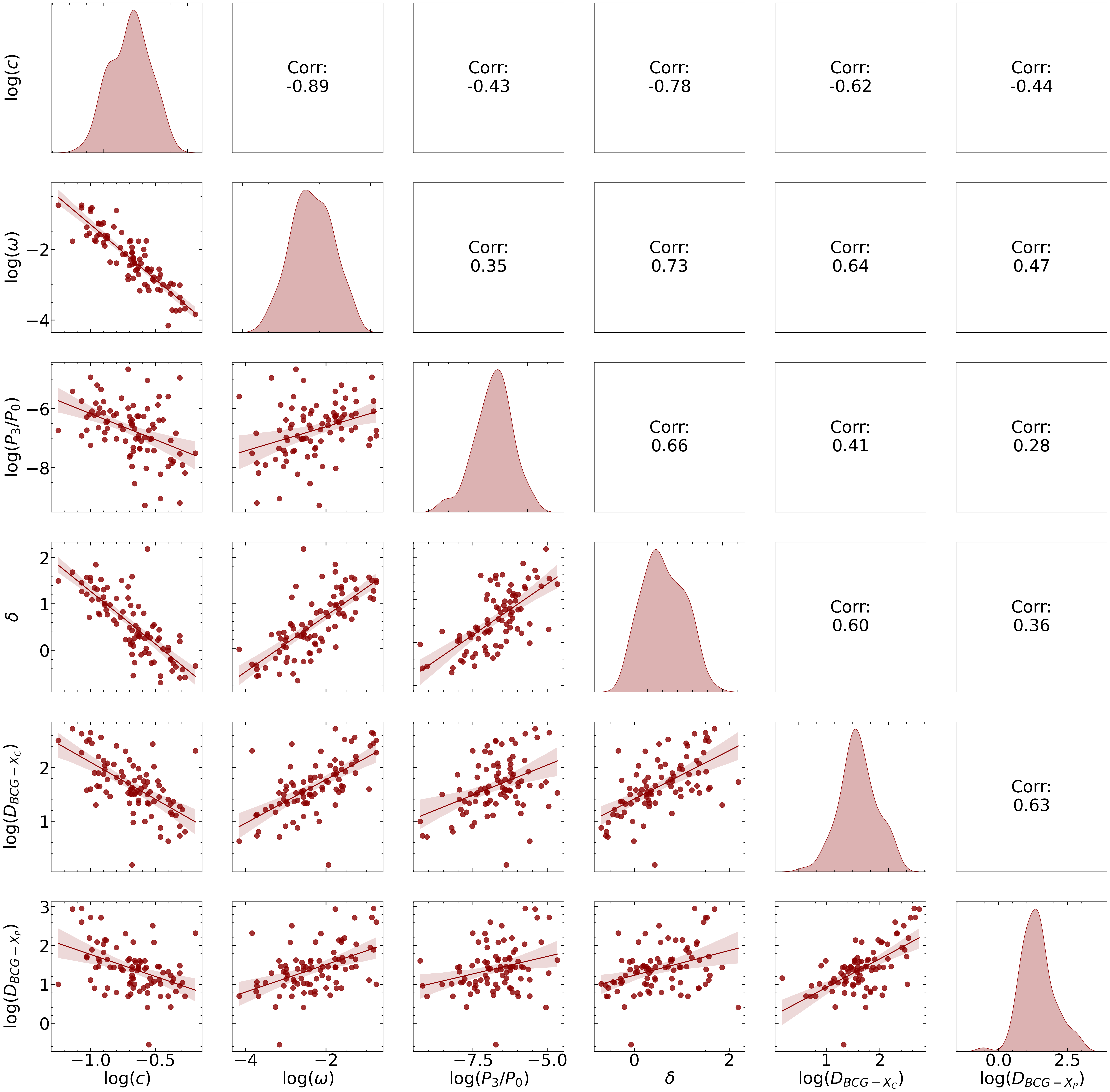}
    \caption{Correlation matrix of the six dynamical state proxies used in this thesis. The diagonal displays the kernel density estimation (KDE) of each variable. Below the diagonal, scatter plots with linear fits and their corresponding confidence intervals are presented for all combinations of these parameters. Above the diagonal, Pearson correlation coefficients associated with each parameter space are shown.}
    \label{fig:dynamical-parameters-correlations}
\end{figure*}

A method that might initially seem effective is to create a Boolean sum of disturbance conditions. Then, one could observe the distribution of this Boolean sum and categorize the sample into relaxed and disturbed clusters, perhaps with an intermediate region. However, as seen in Fig. \ref{fig:dynamical-parameters-correlations}, several parameters are correlated with each other, either positively or negatively. Therefore, this method could potentially give more statistical weight to one parameter over another, leading to biased results.

To address this situation, we take advantage that some clusters in our study sample have dynamical states determined using other methods, such as visual inspection of features in radio and a combination of optical and X-ray observations. We searched for those clusters in the literature, creating a subset that consists of 26 systems with well-defined dynamical states (Table \ref{tab:dynamical-state-literature}), which we used to calibrate our method.

\begin{table}[h]
    \caption{List of galaxy clusters with known dynamical states from literature used to calibrate our method.}
    \label{tab:dynamical-state-literature}
    \centering
    \resizebox{0.4\textwidth}{!}{
    \begin{tabular}{lcr}
    \hline\hline
    Cluster & Dynamical state & Reference\\
    \hline
    Abell 141 & Disturbed & \citet{Cassano+10} \\
    Abell 267 & Relaxed & \citet{Cassano+10} \\
    Abell 384 & Relaxed & \citet{Cuciti+21} \\
    Abell 402 & Relaxed & \citet{Cuciti+21} \\
    Abell 2537 & Relaxed & \citet{Cassano+10} \\
    Abell 2631 & Disturbed & \citet{Cuciti+21} \\
    Abell 2697 & Relaxed & \citet{Cuciti+21} \\
    Abell 2744 & Disturbed & \citet{Cassano+10} \\
    Abell 2813 & Relaxed & \citet{Kesebonye+23} \\
    Abell 2895 & Disturbed & \citet{Kesebonye+23} \\
    Abell 3017 & Intermediate & \citet{Lovisari+17} \\
    Abell 3041 & Disturbed & \citet{Cuciti+21} \\
    Abell 3088 & Relaxed & \citet{Cassano+10} \\
    Abell 3322 & Intermediate & \citet{Lovisari+17} \\
    Abell 3343 & Relaxed & \citet{Kesebonye+23} \\
    Abell 3364 & Relaxed & \citet{Lovisari+17} \\
    Abell 3378 & Relaxed & \citet{Lovisari+17} \\
    Abell 3739 & Relaxed & \citet{Lovisari+17} \\
    Abell 3827 & Relaxed & \citet{Lovisari+17} \\
    Abell S295 & Disturbed & \citet{Kesebonye+23} \\
    Abell S520 & Disturbed & \citet{Kesebonye+23} \\
    Abell S1063 & Relaxed & \citet{Kesebonye+23} \\
    ACT-CLJ0217-5245 & Disturbed & \citet{Kesebonye+23} \\
    RXC J0528.2-2942 & Relaxed & \citet{Kesebonye+23} \\
    RXC J2023.4-5535 & Disturbed & \citet{Kesebonye+23} \\
    SPT-CLJ0232-4421 & Relaxed & \citet{Kesebonye+23} \\
    \hline
    \end{tabular}
    }
\end{table}

\begin{figure}[h]
    \centering
    \includegraphics[width=0.45\textwidth]{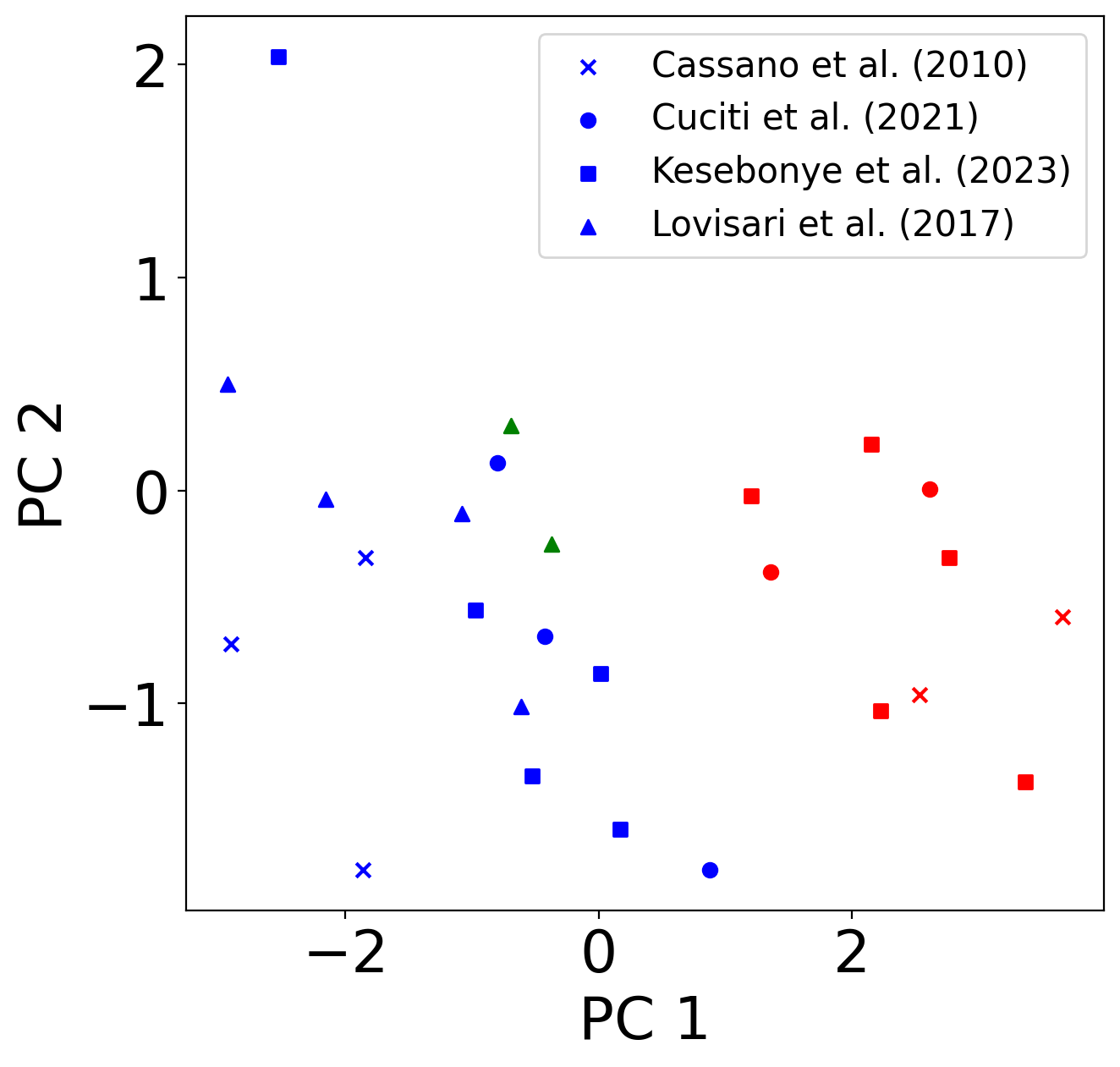}
    \caption{Parameter space of the first two principal components (PC 1 and PC 2) obtained from the six dynamic parameters described in Section \ref{sec:dyn-state-proxies} for the 26 clusters in the sample that have a well-defined dynamical state in the literature. Blue, red, and green symbols represent relaxed, disturbed, and intermediate clusters, respectively. Crosses, dots, squares and triangles correspond to the data extracted from \citet{Cassano+10}, \citet{Cuciti+21}, \citet{Kesebonye+23}, and \citet{Lovisari+17}, respectively.}
    \label{fig:dynamical-parameters-pca}
\end{figure}

Subsequently, we apply Principal Component Analysis (PCA) to this subsample. PCA is a statistical tool that, using a linear transformation, can reduce the dimensionality of a dataset by identifying (linearly) correlated variables and eliminating noise and redundancy in the data. The advantage of this technique is that it can transform a set of possibly related variables into another set of more fundamental independent variables \citep{Hotelling33}. Additionally, if the redundancy is significant and, hence, there is a correlation between the variables, it might be possible to reproduce the original values of the variables with fewer principal components than the original number of variables in the dataset without losing their features.

\begin{table}[h]
    \caption{Normalized statistical weight of the dynamical parameters associated with the first principal component.}
    \label{tab:pca-statistical-weights}
    \centering
    \resizebox{0.22\textwidth}{!}{
    \begin{tabular}{lr}
    \hline\hline
    Parameter & Statistical weight\\
    \hline
    $D_{\text{BCG-X}_{\text{P}}}$ & 0.21\\
    $D_{\text{BCG-X}_{\text{C}}}$ & 0.28\\
    $\delta$ & 0.30\\
    $c$ & -0.30\\
    $\omega$ & 0.30\\
    $P_3/P_0$ & 0.21\\
    \hline
    \end{tabular}
    }
\end{table}

Thus, we apply PCA to our set of variables $\{D_{\text{BCG-X}_{\text{P}}}$, $D_{\text{BCG-X}_{\text{P}}}$,$\delta$, $c$, $\omega$, $P_3/P_0\}$. Fig. \ref{fig:dynamical-parameters-pca} shows that the first principal component (PC 1) can roughly separate the sample into relaxed and disturbed clusters. With this observation, we extract the statistical weights of each dynamical parameter associated with PC 1 (see Table \ref{tab:pca-statistical-weights}). Subsequently, we perform a weighted Boolean sum to determine the dynamical state of the clusters.

\begin{equation}
    \texttt{wbs} = \sum_i w_i \, B_i,
\end{equation}

where $w_i$ is the statistical weight extracted from PCA and $B_i$ is the Boolean value.

\begin{figure}[h]
    \centering
    \includegraphics[width=0.45\textwidth]{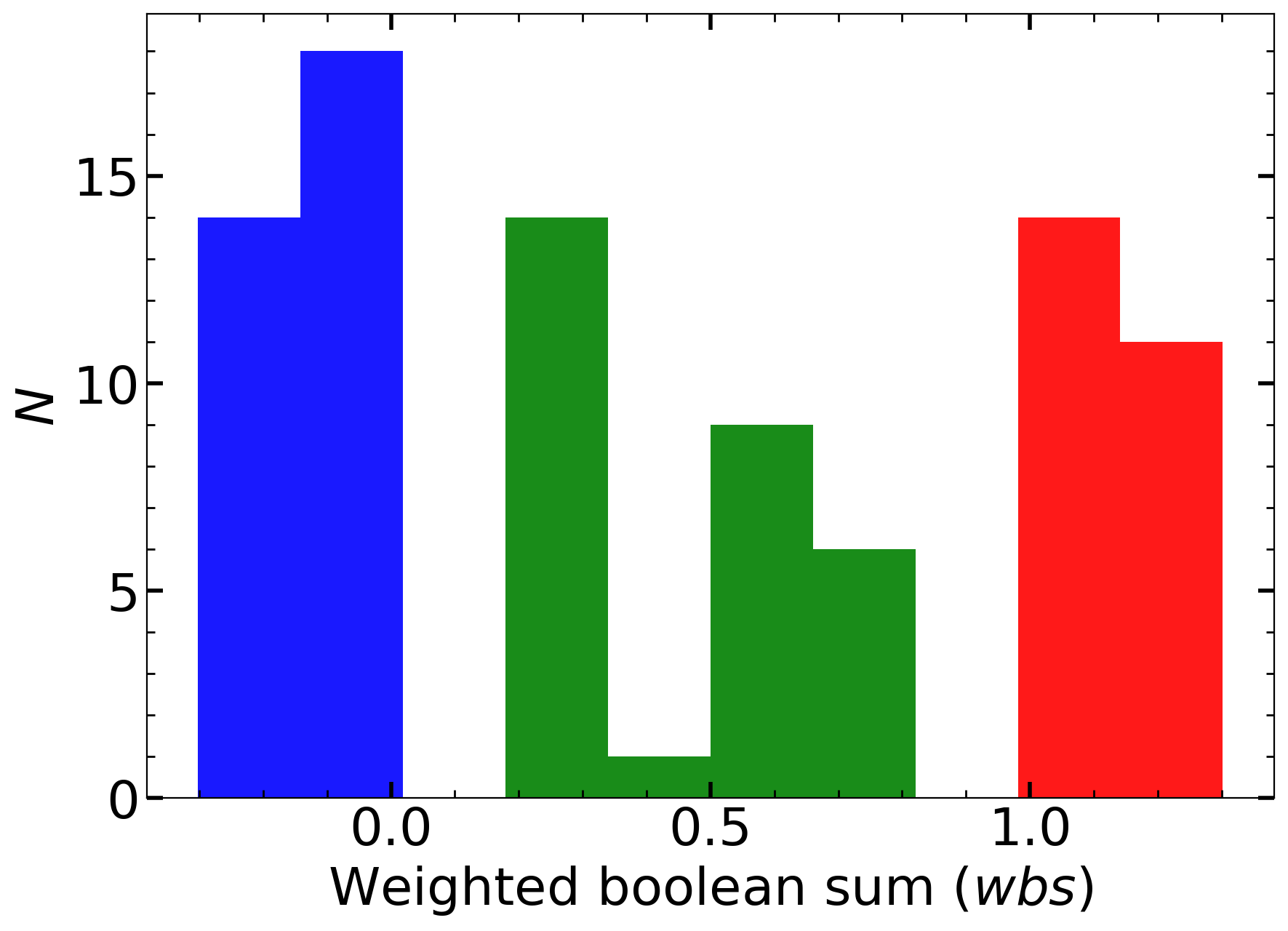}
    \caption{Distribution of the $wbs$ parameter for all the galaxy clusters in the sample. We classify relaxed clusters as systems with $wbs < 0$ (blue bars), disturbed clusters with $wbs > 0$ (red bars), and the remaining clusters in between as an intermediate population of galaxy clusters (green bars).}
    \label{fig:weighted-boolean-sum}
\end{figure}

As observed in the distribution of the weighted Boolean sum (\texttt{wbs}) in Fig. \ref{fig:weighted-boolean-sum}, it is suggestive to divide the sample into three categories: relaxed clusters (\texttt{wbs} $<0$), intermediate clusters ($0 \leq$ \texttt{wbs} $\leq 1$), and disturbed clusters (\texttt{wbs} $>1$). This is the criterion that we use throughout this work.

Using the \texttt{wbs} criterion to estimate the dynamical state of galaxy clusters, we identify 32 relaxed clusters, 30 systems in an intermediate dynamical state, and 25 disturbed galaxy clusters.

\begin{figure}[h]
    \centering
    \includegraphics[width=0.45\textwidth]{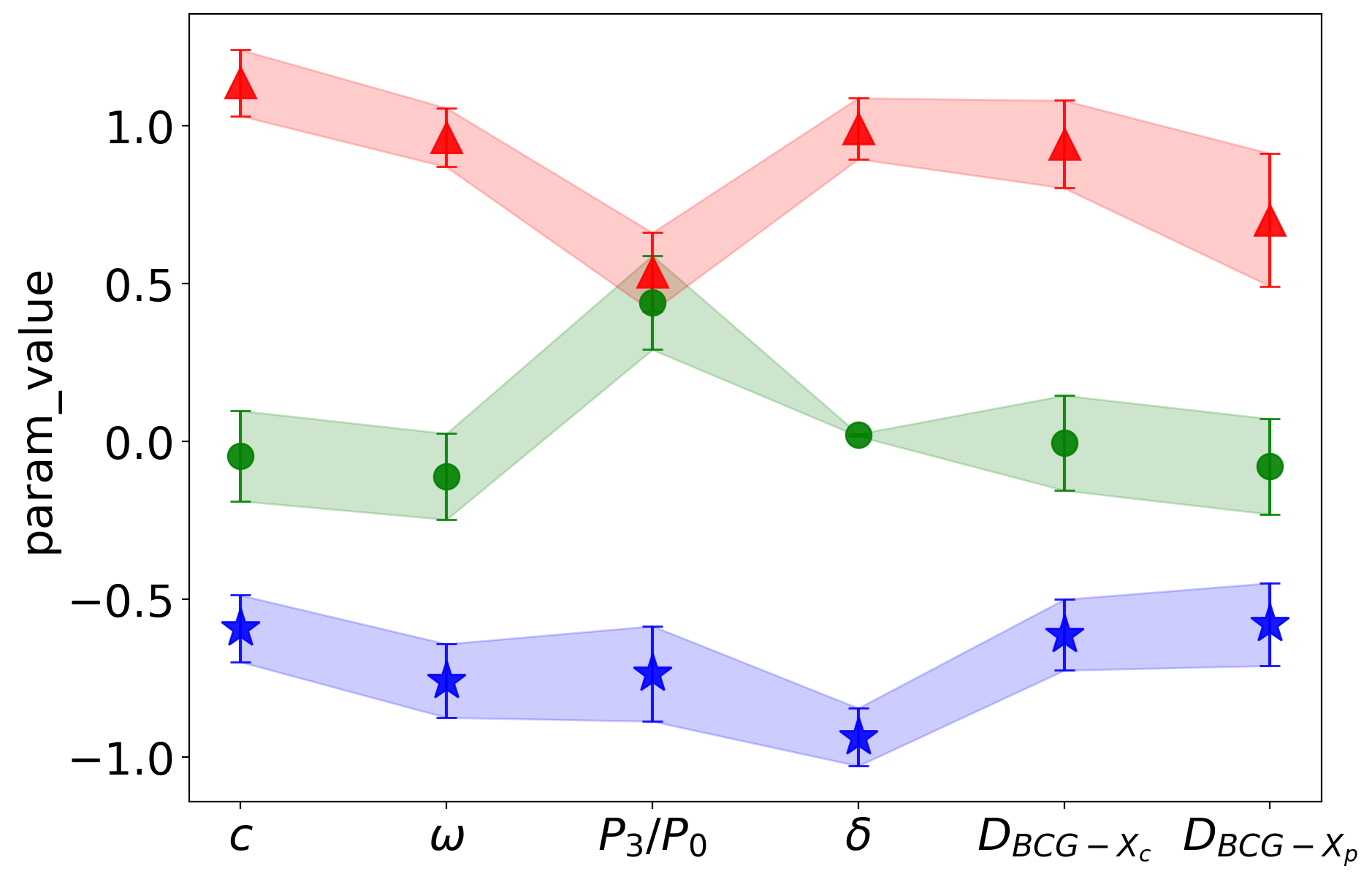}
    \caption{Median values of the six dynamical state proxies for the subsamples of relaxed clusters (blue stars), intermediate clusters (green dots), and disturbed clusters (red triangles). The parameters are standardized, and the radar charts are on the same scale for all clusters for comparison purposes. This scale corresponds to $-3.30 \leq$ \texttt{param\_value} $\leq 2.60$, where \texttt{param\_value} is the dimensionless scaled value of any dynamical parameter. The concentration parameter $c$ is inverted for visualization purposes only. The shaded areas and error bars represent the 1$\sigma$ regions.
}
    \label{fig:median-values-dynamical-parameters}
\end{figure}

In Fig. \ref{fig:radar-charts}, the scaled values for each proxy are shown, with the dynamical states of each system represented by the colors blue, green, and red for the relaxed, intermediate, and disturbed clusters, respectively. The values of each parameter were scaled using the StandardScaler class from the \textsc{scikit-learn} Python package. We note that, except for the concentration parameter, the relaxed clusters generally have values lower than the disturbed clusters for the dynamical state proxies, leaving the intermediate clusters with values precisely in between the two. Furthermore, in Fig. \ref{fig:median-values-dynamical-parameters} we see the median values of each dynamical parameter with their respective standard errors for each dynamical state. We observe a significant difference between relaxed and disturbed clusters, placing the intermediate clusters in a transition zone between them, which gives us confidence in the methodology employed (i.e., PCA). It is essential to mention that the sign of the concentration values has been inverted in Fig. \ref{fig:median-values-dynamical-parameters}. This was done to facilitate visualization, as this is the only proxy where higher values indicate relaxation, and this can lead to misinterpretations when observing an overlapping zone between relaxed and disturbed clusters that do not exist.

\section{Discussion and conclusions}
\label{sec:discussion}

\subsection{BCG selection accuracy}
\label{sec:disc-bcg-selection-accuracy}

We mentioned in Section \ref{sec:bcg-selection} that we employed an automatic method to select the BCG candidates of all the galaxy clusters in the sample. However, we used visual inspection to ensure that the selection was correct, i.e., confirm or reject the BCG candidates selected with the automatic method. In 70\% of the cases, the BCG selected by both methods coincides. For the remaining cases, we attribute the error of the automatic method to three main reasons; the cluster membership method used, the specific physical processes that can occur, and special cases. Specifically, the first is that the method of assigning members using photometric redshifts has a contamination rate that can reach 25\%, and this can cause a redder galaxy than the BCG to be found within the $1\sigma$ region of the best fit of the Red Sequence, but that does not actually belong to the cluster. The second reason is due to the presence of cooling flows in relaxed clusters, where the cold gas directly reaches the BCG, triggering significant star formation and/or nuclear activity \citep[e.g.,][]{Rawle+12}. This causes the BCG to be bluer and, therefore, might not be selected by the automatic method. Regarding special cases, it is possible for a spectroscopic confirmed member galaxy within $R_{200}$ to be brighter than the BCG. An example of this is the galaxy cluster Abell 3827, where four galaxies in the central region are likely to merge into a dominant central galaxy (cD) in the future \citep{carrasco2010}. However, at the time of observation, this has not yet ocurred, and there exists another galaxy that is brighter than any of the four central ones, but it does not meet the other criteria for classification as the BCG (Carrasco et al., in prep).

\begin{figure*}[h]
    \centering
    \includegraphics[width=1\textwidth]{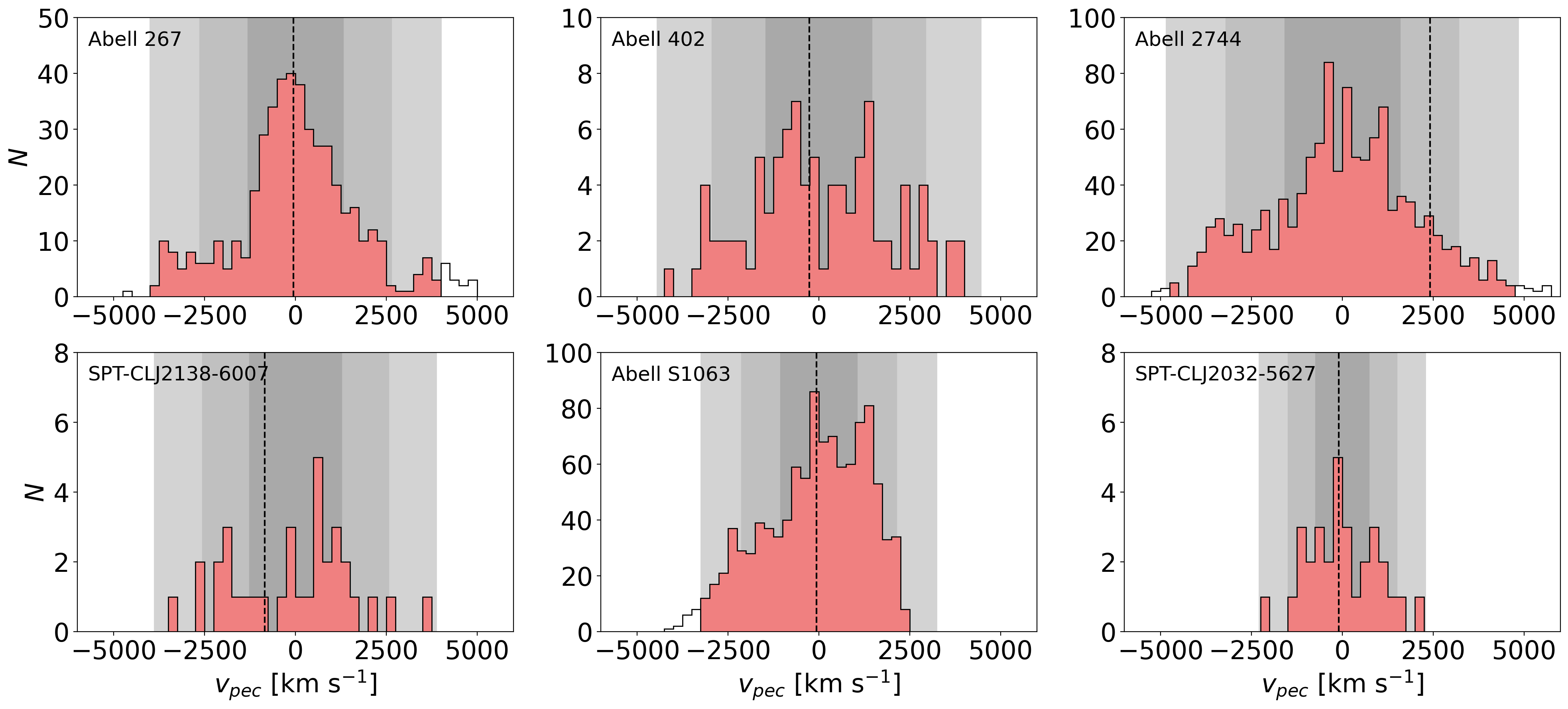}
    \caption{Peculiar velocity distributions of galaxies within 6000 km s$^{-1}$ of six example galaxy clusters. The members are marked with red bars. The black dashed lines indicate the positions of the BCGs selected by visual inspection. The shaded areas represent the 1$\sigma$, 2$\sigma$, and 3$\sigma$ regions from darkest to lightest. The cluster names are indicated in the upper left corners of each panel. The bin sizes are 250 km s$^{-1}$.}
    \label{fig:cluster-peculiar-velocities}
\end{figure*}

Even more importantly, since we measure the accuracy of the automatic method with respect to visual inspection, we must ensure that the selection of the BCG using the latter method is indeed correct. To this end, we have spectroscopic information from the literature (Table \ref{tab:spectroscopic-catalogs}). In Fig. \ref{fig:cluster-peculiar-velocities}, we present as an example the peculiar velocities of six galaxy clusters from the sample within the range of $\pm 6000$ km s$^{-1}$, and we observe that all the BCGs selected visually indeed belong to the cluster. This occurs for the 30 systems for which we have spectroscopic information for the BCG selected by visual inspection, which gives us confidence in the method.

It is important to clarify that the accuracy of the automatic method does not have any effect on the results of this work, because, as we previously said, the candidates for the BCG were then confirmed/rejected by visual inspection and spectroscopic redshifts when available. We provide the accuracy of the method only as a reference to take into account what can be improved in future works.

\subsection{Dynamical parameters}

Although classifying dynamical states based on a single parameter can help to distinguish the most extreme cases, a proper combination of these indices can be more robust. To quantify this, we compared our results with those obtained by \citet{Yuan&Han20} and \citet{YuanHan&Wen22}. For disturbed clusters, we find consistent results between our classification system and that used by \citet{Yuan&Han20} and \citet{YuanHan&Wen22} ($\delta > 0$). However, we note a discrepancy for the relaxed clusters. Of the 32 relaxed galaxy clusters in our sample, 13 do not fall into this category according to the \citet{Yuan&Han20} and \citet{YuanHan&Wen22} criteria ($\delta < 0$)\footnote{This subsample of clusters corresponds to Abell 122, Abell 223, Abell 3322, Abell 3364, Abell 3718, Abell 3783, MCXC J0528.9-3927, PLCKESZG256.4-65, PLCKG334.8-38.0A, RXC J0439.2-4600, RXC J0528.2-2942, SPT-CLJ0106-5943, and SPT-CLJ2130-6458.}. This fraction corresponds to 40\%, and since this is a rather high value, we have individually analyzed these 13 systems. They all have a $wbs$ value of -0.0057, which is almost at the threshold of our separation between relaxed and intermediate clusters. However, upon inspection, only Abell 3322, Abell 3783, and MCXCJ0528.9-3927 exhibit interaction features typical of an unrelaxed cluster (see Figs. in \url{https://zenodo.org/records/14241676}). This indicates the need for an intermediate systems category rather than opting for a binary classification between relaxed and disturbed clusters, and our combined use of dynamical state proxies allows us to do this robustly.

The binary separation between relaxed and disturbed clusters arises from the bimodalities found in the distributions of these dynamical parameters. However, in some cases, the bimodalities may be absent, making the classification task more challenging \citep[e.g.,][]{Campitiello+22}. The use of two dynamical state proxies together in a two-dimensional parameter space has been shown to aid in this classification task \citep[e.g.,][]{Cassano+10,Cassano+13}. However, even more robust results can be obtained by combining multiple dynamical parameters, such as with the $M$ statistic \citep[e.g.,][]{Rasia+13,Lovisari+17}. In our case, we have chosen to use PCA with our set of six proxies, finding that the first principal component can effectively separate the dynamical states of the clusters, which is consistent with \citet{Campitiello+22}.

\subsection{Classification of dynamical states}

From our total sample of 87 galaxy clusters, we classified 32 ($\sim 36\%$) as relaxed clusters, 30 ($\sim 34\%$) as systems with an intermediate dynamical state, and 25 ($\sim 29\%$) as disturbed clusters of galaxies. Thus, we identified a fraction of $\sim 63\%$ of galaxy clusters in an unrelaxed state (intermediate and disturbed), which is consistent with the expected range of $30-80\%$ according to previous studies \citep[e.g.,][]{DresslerShectman88,Santos+08,FakhouriMa10,WenHan13,Yuan&Han20,YuanHan&Wen22}. This sample will then be our starting point in Paper II to study the effect of the dynamical state of galaxy clusters on their populations, focusing on the physical and structural properties of member galaxies.

\section*{Data availability}

We provide three figures via Zenodo at \url{https://zenodo.org/records/14241676}, showing the contours of the X-ray surface brightness distribution, the distribution of red sequence galaxies, and the positions of the BGGs, X-ray peaks, and X-ray centroids.

\begin{acknowledgements}
      JLNC \& SVA  acknowledges the financial support of DIDULS/ULS, through the Proyecto Apoyo de Tesis de Postgrado N° PTE2353858. ERC acknowledges the support of the International Gemini Observatory, a program of NSF NOIRLab, which is managed by the Association of Universities for Research in Astronomy (AURA) under a cooperative agreement with the U.S. National Science Foundation, on behalf of the Gemini partnership of Argentina, Brazil, Canada, Chile, the Republic of Korea, and the United States of America. We wish to thank the anonymous referee for a constructive report that helped us improve our manuscript.
\end{acknowledgements}

%
%

\bibliographystyle{aa}
\bibliography{bibliography}

\begin{appendix}

\onecolumn
\begin{landscape}

\section{Full cluster catalog}

\begin{scriptsize}
\begin{longtable}{lcccccccccccccr}
\caption{\label{kstars} Relevant information about the full sample of galaxy clusters. Columns: (1) galaxy cluster name; (2-3) right ascension and declination in J2000; (4) redshift; (5) characteristic magnitude in the $r$-band; (6) $R_{200}$ in Mpc; (7) $M_{200}$ in $10^{14}M/M_{\odot}$ units; (8) cluster member galaxies using photometric redshifts; (9) concentration; (10) centroid shift; (11) power ratio; (12) morphological parameter; (13) BCG/X-ray peak offset in kpc; (14) BCG/X-ray centroid offset in kpc; (15) dynamical state.}\\
\hline\hline
Name & RA & Dec & $z$ & $m_r^*$ & $R_{200}$ & $M_{200}$ & $N$ & $\log(c)$ & $\log(\omega)$ & $\log(P_3/P_0)$ & $\delta$ & $D_{BCG-X_P}$ & $D_{BCG-X_C}$ & Dynamical state  \\
(1) & (2) & (3) & (4) & (5) & (6) & (7) & (8) & (9) & (10) & (11) & (12) & (13) & (14) & (15) \\
\hline
\endfirsthead
\caption{continued.}\\
\hline\hline
Name & RA & Dec & $z$ & $m_r^*$ & $R_{200}$ & $M_{200}$ & $N$ & $\log(c)$ & $\log(\omega)$ & $\log(P_3/P_0)$ & $\delta$ & $D_{BCG-X_P}$ & $D_{BCG-X_C}$ & Dynamical state \\
(1) & (2) & (3) & (4) & (5) & (6) & (7) & (8) & (9) & (10) & (11) & (12) & (13) & (14) & (15) \\
\hline
\endhead
\hline
\endfoot
Abell 2715 & 0.6792 & -34.6622 & 0.116 & 16.89 & 1.24 & 3.08 & 39 & -0.76 ± 0.01 & -1.52 ± 0.01 & -5.64 ± 0.03 & 1.51 ± 0.01 & 24 & 83 & Disturbed \\
Abell 2697 & 0.8004 & -6.0937 & 0.248 & 18.75 & 1.56 & 6.12 & 154 & -0.66 ± 0.02 & -2.38 ± 0.02 & -6.67 ± 0.13 & -0.27 ± 0.01 & 40 & 57 & Relaxed \\
Abell 2744 & 3.5800 & -30.3922 & 0.308 & 19.31 & 1.95 & 11.91 & 146 & -1.01 ± 0.03 & -1.54 ± 0.01 & -6.19 ± 0.11 & 0.71 ± 0.01 & 155 & 336 & Disturbed \\
CL0019.6+0336 & 4.9101 & 3.6022 & 0.269 & 18.95 & 1.81 & 9.48 & 148 & -0.70 ± 0.01 & -1.78 ± 0.01 & -6.26 ± 0.09 & 0.49 ± 0.01 & 45 & 139 & Disturbed \\
Abell S67 & 10.2523 & -44.4871 & 0.324 & 19.44 & 1.79 & 9.22 & 125 & -0.71 ± 0.02 & -2.75 ± 0.05 & -4.66 ± 0.04 & 1.37 ± 0.01 & 41 & 33 & Intermediate \\
Abell 2811 & 10.5310 & -28.5360 & 0.108 & 16.72 & 1.42 & 4.59 & 134 & -0.61 ± 0.01 & -2.39 ± 0.01 & -7.10 ± 0.09 & -0.10 ± 0.01 & 37 & 40 & Relaxed \\
Abell 2813 & 10.8532 & -20.6246 & 0.292 & 19.17 & 1.95 & 11.83 & 99 & -0.67 ± 0.02 & -2.26 ± 0.04 & -7.32 ± 0.33 & 0.09 ± 0.01 & 122 & 108 & Intermediate \\
Abell S84 & 12.3461 & -29.5203 & 0.108 & 16.72 & 1.53 & 5.69 & 107 & -0.46 ± 0.01 & -3.16 ± 0.05 & -9.05 ± 0.51 & -0.56 ± 0.01 & 4 & 5 & Relaxed \\
Abell 122 & 14.3450 & -26.2826 & 0.114 & 16.84 & 1.37 & 4.07 & 23 & -0.66 ± 0.01 & -2.40 ± 0.01 & -8.54 ± 0.33 & 0.26 ± 0.01 & 10 & 31 & Relaxed \\
WHL J010455.4+000336 & 16.2303 & 0.0605 & 0.277 & 19.02 & 1.76 & 8.73 & 172 & -0.87 ± 0.01 & -1.62 ± 0.01 & -6.67 ± 0.04 & 1.36 ± 0.01 & 8 & 121 & Disturbed \\
Abell 141 & 16.3870 & -24.6453 & 0.230 & 18.55 & 1.85 & 10.12 & 40 & -0.92 ± 0.02 & -1.29 ± 0.01 & -5.34 ± 0.06 & 1.53 ± 0.01 & 517 & 368 & Disturbed \\
SPT-CLJ0106-5943 & 16.6184 & -59.7206 & 0.348 & 19.63 & 1.49 & 5.24 & 186 & -0.54 ± 0.01 & -2.62 ± 0.01 & -7.03 ± 0.03 & 0.27 ± 0.01 & 14 & 35 & Relaxed \\
Z348 & 16.7062 & 1.0558 & 0.251 & 18.78 & 1.27 & 3.24 & 72 & -0.19 ± 0.01 & -3.84 ± 0.01 & -7.51 ± 0.02 & -0.34 ± 0.01 & 207 & 207 & Intermediate \\
RXC J0117.8-5455 & 19.4642 & -54.9224 & 0.251 & 18.77 & 1.16 & 2.47 & 119 & -0.40 ± 0.01 & -4.16 ± 0.01 & -5.59 ± 0.02 & -0.01 ± 0.01 & 4 & 4 & Relaxed \\
Abell 2895 & 19.5476 & -26.9669 & 0.227 & 18.52 & 1.56 & 6.04 & 74 & -0.87 ± 0.02 & -1.68 ± 0.01 & -6.42 ± 0.09 & 0.77 ± 0.01 & 43 & 60 & Disturbed \\
PSZ1G295.60-51.95 & 23.3620 & -64.5695 & 0.333 & 19.51 & 1.50 & 5.41 & 66 & -1.14 ± 0.03 & -1.77 ± 0.08 & -5.41 ± 0.08 & 1.69 ± 0.01 & 864 & 534 & Disturbed \\
Abell 222 & 24.3920 & -12.9918 & 0.213 & 18.35 & 1.73 & 8.26 & 111 & -0.89 ± 0.02 & -1.90 ± 0.02 & -6.81 ± 0.24 & 0.98 ± 0.01 & 5 & 110 & Disturbed \\
Abell 223 & 24.4832 & -12.8216 & 0.207 & 18.28 & 1.86 & 10.33 & 42 & -0.63 ± 0.01 & -2.71 ± 0.02 & -6.92 ± 0.09 & 0.30 ± 0.01 & 25 & 24 & Relaxed \\
RXC J0138.0-2155 & 24.5161 & -21.9260 & 0.338 & 19.55 & 1.65 & 7.20 & 182 & -0.29 ± 0.01 & -3.51 ± 0.01 & -7.91 ± 0.15 & -0.42 ± 0.01 & 11 & 16 & Relaxed \\
Abell 2941 & 26.2360 & -53.0185 & 0.118 & 16.93 & 1.54 & 5.85 & 118 & -0.89 ± 0.01 & -1.61 ± 0.01 & -7.05 ± 0.10 & 1.01 ± 0.01 & 139 & 92 & Disturbed \\
ZGX J015223-140420 & 28.0901 & -14.0881 & 0.330 & 19.49 & 1.27 & 3.28 & 133 & -0.90 ± 0.02 & -1.81 ± 0.04 & -6.14 ± 0.16 & 1.15 ± 0.01 & 66 & 75 & Disturbed \\
MACS J0152.5-2852 & 28.1443 & -28.8939 & 0.341 & 19.58 & 1.68 & 7.63 & 162 & -0.52 ± 0.03 & -2.85 ± 0.07 & -6.53 ± 0.05 & 0.52 ± 0.01 & 322 & 276 & Intermediate \\
Abell 267 & 28.1759 & 1.0099 & 0.233 & 18.58 & 1.44 & 4.72 & 113 & -0.58 ± 0.01 & -2.17 ± 0.01 & -9.28 ± 0.34 & -0.04 ± 0.01 & 40 & 9 & Relaxed \\
RXC J0153.5-0118 & 28.3934 & -1.3022 & 0.244 & 18.70 & 1.33 & 3.72 & 92 & -0.80 ± 0.02 & -2.39 ± 0.03 & -6.17 ± 0.13 & 0.57 ± 0.01 & 6 & 66 & Intermediate \\
Abell 286 & 29.6108 & -1.7782 & 0.160 & 17.65 & 1.38 & 4.21 & 95 & -0.77 ± 0.02 & -1.95 ± 0.02 & -6.57 ± 0.09 & 0.71 ± 0.01 & 28 & 53 & Intermediate \\
GMBCG J030.1+00.7 & 30.1278 & 0.7409 & 0.345 & 19.61 & 1.07 & 1.98 & 49 & -0.31 ± 0.01 & -3.36 ± 0.01 & -4.95 ± 0.03 & 0.31 ± 0.01 & 19 & 18 & Intermediate \\
WHL J020046-064230 & 30.1921 & -6.7081 & 0.338 & 19.55 & 1.24 & 3.08 & 86 & -0.71 ± 0.02 & -2.85 ± 0.06 & -5.98 ± 0.16 & 1.13 ± 0.01 & 4 & 51 & Intermediate \\
Abell 291 & 30.4300 & -2.1973 & 0.197 & 18.16 & 1.32 & 3.69 & 97 & -0.31 ± 0.01 & -3.72 ± 0.26 & -9.20 ± 0.56 & -0.60 ± 0.01 & 9 & 5 & Relaxed \\
ACT-CLJ0217-5245 & 34.2794 & -52.7494 & 0.343 & 19.59 & 1.45 & 4.84 & 58 & -1.07 ± 0.01 & -0.75 ± 0.05 & -6.92 ± 0.41 & 1.46 ± 0.01 & 399 & 190 & Disturbed \\
RXC J0220.9-3829 & 35.2359 & -38.4808 & 0.228 & 18.53 & 1.34 & 3.84 & 67 & -0.37 ± 0.01 & -3.72 ± 0.01 & -7.84 ± 0.06 & -0.45 ± 0.01 & 11 & 13 & Relaxed \\
PLCKESZG256.4-65 & 36.3556 & -42.0147 & 0.220 & 18.44 & 1.51 & 5.49 & 164 & -0.58 ± 0.01 & -2.99 ± 0.05 & -6.97 ± 0.18 & 0.25 ± 0.01 & 25 & 45 & Relaxed \\
WHL J022544-031233 & 36.4286 & -3.2094 & 0.141 & 17.35 & 1.30 & 3.52 & 53 & -0.99 ± 0.01 & -0.82 ± 0.01 & -6.08 ± 0.02 & 1.09 ± 0.01 & 77 & 286 & Disturbed \\
Abell 3017 & 36.4714 & -41.9167 & 0.220 & 18.43 & 1.66 & 7.30 & 73 & -0.48 ± 0.02 & -3.13 ± 0.08 & -5.87 ± 0.10 & 0.22 ± 0.01 & 28 & 56 & Intermediate \\
Abell 362 & 37.9199 & -4.8862 & 0.184 & 17.99 & 1.75 & 8.59 & 54 & -0.94 ± 0.01 & -1.27 ± 0.01 & -6.64 ± 0.14 & 0.80 ± 0.01 & 44 & 79 & Disturbed \\
SPT-CLJ0232-4421 & 38.0789 & -44.3469 & 0.284 & 19.09 & 1.73 & 8.30 & 133 & -0.51 ± 0.01 & -2.89 ± 0.06 & -6.65 ± 0.15 & 0.55 ± 0.01 & 119 & 104 & Intermediate \\
ACT-CLJ0235-5121 & 38.9377 & -51.3530 & 0.278 & 19.04 & 1.56 & 6.01 & 137 & -0.82 ± 0.01 & -1.36 ± 0.01 & -6.23 ± 0.10 & 0.65 ± 0.01 & 27 & 115 & Disturbed \\
Abell 368 & 39.3658 & -26.5082 & 0.220 & 18.44 & 1.52 & 5.64 & 110 & -0.38 ± 0.01 & -3.26 ± 0.01 & -7.72 ± 0.04 & -0.21 ± 0.01 & 2 & 14 & Relaxed \\
Abell 3038 & 39.4993 & -52.4136 & 0.135 & 17.24 & 1.29 & 3.43 & 138 & -0.59 ± 0.01 & -2.00 ± 0.01 & -6.26 ± 0.02 & 0.80 ± 0.01 & 8 & 56 & Intermediate \\
WHL J023941-012812 & 39.9306 & -1.4682 & 0.325 & 19.44 & 1.31 & 3.57 & 129 & -0.95 ± 0.03 & -1.62 ± 0.04 & -5.20 ± 0.05 & 1.32 ± 0.01 & 122 & 45 & Disturbed \\
Abell 3041 & 40.3412 & -28.6540 & 0.235 & 18.61 & 1.90 & 10.93 & 110 & -0.90 ± 0.02 & -1.25 ± 0.01 & -5.76 ± 0.05 & 1.31 ± 0.01 & 107 & 136 & Disturbed \\
MCXC J0244.1-2611 & 41.0526 & -26.1748 & 0.136 & 17.26 & 1.57 & 6.15 & 81 & -0.97 ± 0.01 & -1.74 ± 0.02 & -6.22 ± 0.08 & 1.35 ± 0.01 & 7 & 79 & Disturbed \\
Abell S295 & 41.3625 & -53.0291 & 0.300 & 19.24 & 1.48 & 5.19 & 198 & -0.80 ± 0.01 & -0.90 ± 0.01 & -7.02 ± 0.06 & 1.12 ± 0.01 & 88 & 256 & Disturbed \\
Abell 3048 & 41.4935 & -20.4882 & 0.309 & 19.31 & 1.37 & 4.07 & 65 & -0.64 ± 0.02 & -2.57 ± 0.04 & -6.25 ± 0.14 & 0.35 ± 0.01 & 22 & 37 & Intermediate \\
PLCKG205.0-63.0 & 41.6076 & -20.5556 & 0.310 & 19.32 & 1.71 & 7.95 & 97 & -1.03 ± 0.02 & -1.60 ± 0.02 & -6.28 ± 0.24 & 1.21 ± 0.01 & 48 & 37 & Disturbed \\
Abell 383 & 42.0146 & -3.5291 & 0.190 & 18.07 & 1.50 & 5.39 & 45 & -0.27 ± 0.01 & -3.68 ± 0.03 & -8.18 ± 0.04 & -0.59 ± 0.01 & 4 & 6 & Relaxed \\
Abell 384 & 42.0498 & -2.2765 & 0.236 & 18.61 & 1.63 & 6.87 & 119 & -0.73 ± 0.02 & -2.13 ± 0.02 & -7.10 ± 0.24 & 0.23 ± 0.01 & 203 & 202 & Intermediate \\
Abell 402 & 44.4229 & -22.1556 & 0.322 & 19.43 & 1.96 & 11.94 & 26 & -0.47 ± 0.01 & -2.56 ± 0.01 & -6.39 ± 0.01 & 0.26 ± 0.01 & 28 & 28 & Intermediate \\
WHL J025932+001354 & 44.8850 & 0.2311 & 0.201 & 18.21 & 1.24 & 3.01 & 66 & -0.69 ± 0.02 & -1.75 ± 0.03 & -6.43 ± 0.12 & 0.97 ± 0.01 & 11 & 39 & Intermediate \\
Abell 3088 & 46.7590 & -28.6666 & 0.253 & 18.8 & 1.57 & 6.22 & 77 & -0.46 ± 0.02 & -2.71 ± 0.03 & -8.22 ± 0.56 & -0.70 ± 0.01 & 10 & 7 & Relaxed \\
MCXC J0320.6-4311 & 50.1552 & -43.1972 & 0.149 & 17.48 & 1.38 & 4.20 & 47 & -0.67 ± 0.01 & -2.24 ± 0.04 & -5.89 ± 0.03 & 0.79 ± 0.01 & 13 & 13 & Intermediate \\
MCXC J0336.8-2804 & 54.2103 & -28.0829 & 0.105 & 16.66 & 1.53 & 5.69 & 35 & -0.85 ± 0.01 & -2.36 ± 0.02 & -6.07 ± 0.04 & 0.98 ± 0.01 & 27 & 28 & Intermediate \\
SPT-CLJ0348-4514 & 57.0706 & -45.2492 & 0.325 & 19.45 & 1.70 & 7.87 & 151 & -0.68 ± 0.01 & -1.94 ± 0.01 & -6.38 ± 0.03 & 0.43 ± 0.01 & 14 & 1 & Intermediate \\
Abell 3213 & 61.0708 & -27.0935 & 0.250 & 18.76 & 1.77 & 8.84 & 149 & -0.56 ± 0.01 & -2.56 ± 0.01 & -5.04 ± 0.02 & 2.19 ± 0.01 & 2 & 53 & Intermediate \\
WHY J040650-565840 & 61.7102 & -56.9786 & 0.226 & 18.51 & 1.18 & 2.65 & 149 & -0.68 ± 0.02 & -2.10 ± 0.07 & -6.51 ± 0.12 & 1.59 ± 0.01 & 16 & 42 & Intermediate \\
RXC J0439.2-4600 & 69.8091 & -46.0146 & 0.340 & 19.57 & 1.26 & 3.18 & 26 & -0.40 ± 0.01 & -3.06 ± 0.01 & -7.96 ± 0.12 & 0.06 ± 0.01 & 21 & 25 & Relaxed \\
Abell S506 & 75.2857 & -24.4208 & 0.320 & 19.41 & 1.52 & 5.59 & 41 & -0.93 ± 0.02 & -1.58 ± 0.02 & -5.98 ± 0.11 & 0.77 ± 0.01 & 38 & 44 & Intermediate \\
Abell 3322 & 77.5695 & -45.3200 & 0.200 & 18.20 & 1.64 & 7.06 & 123 & -0.65 ± 0.02 & -2.60 ± 0.05 & -7.02 ± 0.11 & 0.53 ± 0.01 & 36 & 33 & Relaxed \\
Abell S520 & 79.1508 & -54.5021 & 0.295 & 19.19 & 1.77 & 8.89 & 83 & -1.03 ± 0.02 & -1.37 ± 0.01 & -6.73 ± 0.45 & 1.57 ± 0.01 & 51 & 312 & Disturbed \\
SPT-CLJ0522-4818 & 80.5664 & -48.3056 & 0.296 & 19.20 & 1.33 & 3.75 & 137 & -0.44 ± 0.01 & -3.00 ± 0.05 & -6.38 ± 0.02 & 0.03 ± 0.01 & 20 & 38 & Intermediate \\
Abell 3343 & 81.4546 & -47.2528 & 0.191 & 18.09 & 1.52 & 5.62 & 47 & -0.55 ± 0.01 & -3.16 ± 0.01 & -6.73 ± 0.01 & -0.07 ± 0.01 & 0 & 21 & Relaxed \\
RXC J0528.2-2942 & 82.0622 & -29.7215 & 0.158 & 17.62 & 1.39 & 4.31 & 122 & -0.62 ± 0.01 & -3.17 ± 0.03 & -7.35 ± 0.04 & 0.34 ± 0.01 & 40 & 33 & Relaxed \\
MCXC J0528.9-3927 & 82.2215 & -39.4719 & 0.263 & 18.89 & 1.55 & 5.92 & 117 & -0.62 ± 0.01 & -2.28 ± 0.01 & -7.36 ± 0.36 & 0.29 ± 0.01 & 6 & 65 & Relaxed \\
RXC J0532.9-3701 & 83.2323 & -37.0273 & 0.275 & 19.01 & 1.88 & 10.58 & 82 & -0.53 ± 0.01 & -2.56 ± 0.03 & -7.02 ± 0.03 & -0.28 ± 0.01 & 10 & 12 & Relaxed \\
Abell 3364 & 86.9093 & -31.8709 & 0.148 & 17.47 & 1.61 & 6.65 & 50 & -0.68 ± 0.01 & -2.09 ± 0.02 & -7.96 ± 0.28 & 0.16 ± 0.01 & 28 & 38 & Relaxed \\
Abell 3378 & 91.4753 & -35.3029 & 0.141 & 17.35 & 1.33 & 3.78 & 76 & -0.34 ± 0.01 & -3.74 ± 0.01 & -6.83 ± 0.01 & -0.36 ± 0.01 & 6 & 12 & Relaxed \\
Abell S579 & 94.1357 & -39.7998 & 0.152 & 17.53 & 1.30 & 3.47 & 210 & -0.75 ± 0.01 & -2.14 ± 0.03 & -6.71 ± 0.12 & 0.83 ± 0.01 & 23 & 69 & Intermediate \\
RXC J2011.3-5725 & 302.8642 & -57.4201 & 0.279 & 19.04 & 1.15 & 2.40 & 33 & -0.38 ± 0.01 & -2.96 ± 0.11 & -7.08 ± 0.12 & -0.19 ± 0.01 & 16 & 20 & Relaxed \\
RXC J2023.4-5535 & 305.8407 & -55.5967 & 0.232 & 18.57 & 2.01 & 12.98 & 130 & -0.94 ± 0.01 & -1.26 ± 0.01 & -6.16 ± 0.06 & 1.10 ± 0.01 & 28 & 160 & Disturbed \\
SPTCL J2031-4037 & 307.9709 & -40.6238 & 0.342 & 19.58 & 1.50 & 5.36 & 84 & -0.63 ± 0.01 & -1.77 ± 0.01 & -6.25 ± 0.01 & 0.95 ± 0.01 & 27 & 95 & Intermediate \\
SPTCL J2032-5627 & 308.1194 & -56.4836 & 0.284 & 19.09 & 1.52 & 5.62 & 203 & -1.07 ± 0.02 & -0.82 ± 0.01 & -5.74 ± 0.13 & 1.27 ± 0.01 & 896 & 432 & Disturbed \\
PLCKG334.8-38.0A & 313.0701 & -61.2088 & 0.350 & 19.65 & 1.12 & 2.24 & 85 & -0.56 ± 0.01 & -3.05 ± 0.04 & -7.41 ± 0.44 & 0.36 ± 0.01 & 9 & 31 & Relaxed \\
PLCKG334.8-38.0B & 313.2791 & -61.1880 & 0.350 & 19.65 & 1.30 & 3.55 & 74 & -1.00 ± 0.01 & -0.93 ± 0.03 & -7.24 ± 0.41 & 1.57 ± 0.01 & 12 & 39 & Intermediate \\
Abell 3718 & 313.9832 & -54.9268 & 0.139 & 17.31 & 1.52 & 5.64 & 82 & -0.31 ± 0.01 & -3.01 ± 0.05 & -7.53 ± 0.10 & 0.19 ± 0.01 & 30 & 8 & Relaxed \\
Abell 3739 & 316.0792 & -41.3459 & 0.165 & 17.73 & 1.50 & 5.44 & 74 & -0.70 ± 0.01 & -2.46 ± 0.05 & -7.63 ± 0.41 & 0.22 ± 0.01 & 10 & 21 & Intermediate \\
RM J2118.8+0033 & 319.7382 & 0.5479 & 0.270 & 18.96 & 1.80 & 9.24 & 54 & -1.00 ± 0.01 & -0.86 ± 0.01 & -4.93 ± 0.02 & 1.50 ± 0.28 & 526 & 452 & Disturbed \\
RBS1748 & 322.4171 & 0.0883 & 0.235 & 18.6 & 1.51 & 5.52 & 141 & -0.36 ± 0.01 & -2.99 ± 0.03 & -7.78 ± 0.19 & -0.29 ± 0.01 & 13 & 23 & Relaxed \\
WHL J213004-002108 & 322.5165 & -0.3526 & 0.243 & 18.69 & 1.42 & 4.59 & 172 & -0.56 ± 0.02 & -2.55 ± 0.02 & -6.00 ± 0.06 & 0.32 ± 0.01 & 6 & 21 & Intermediate \\
WHL J213027-000024 & 322.6115 & -0.0096 & 0.143 & 17.38 & 1.28 & 3.34 & 42 & -0.96 ± 0.02 & -1.77 ± 0.02 & -5.81 ± 0.05 & 1.85 ± 0.01 & 27 & 20 & Intermediate \\
SPT-CLJ2130-6458 & 322.7347 & -64.9796 & 0.316 & 19.37 & 1.28 & 3.36 & 40 & -0.63 ± 0.01 & -2.50 ± 0.05 & -7.00 ± 0.19 & 0.44 ± 0.01 & 28 & 21 & Relaxed \\
Abell 3783 & 323.5031 & -42.6477 & 0.196 & 18.14 & 1.95 & 11.85 & 86 & -0.69 ± 0.01 & -2.28 ± 0.04 & -7.58 ± 0.38 & 0.12 ± 0.01 & 5 & 32 & Relaxed \\
SPT-CLJ2138-6007 & 324.5053 & -60.1328 & 0.319 & 19.40 & 1.80 & 9.38 & 40 & -0.68 ± 0.02 & -2.42 ± 0.04 & -6.56 ± 0.20 & -0.09 ± 0.01 & 23 & 31 & Relaxed \\
Abell 3827 & 330.4739 & -59.9464 & 0.099 & 16.51 & 1.62 & 6.77 & 136 & -0.67 ± 0.01 & -2.82 ± 0.03 & -7.45 ± 0.26 & -0.55 ± 0.01 & 8 & 9 & Relaxed \\
Abell 3830 & 330.9509 & -61.6008 & 0.211 & 18.33 & 1.26 & 3.23 & 61 & -0.47 ± 0.02 & -2.81 ± 0.03 & -5.59 ± 0.02 & 0.44 ± 0.01 & 13 & 46 & Intermediate \\
Abell S1063 & 342.1853 & -44.5311 & 0.348 & 19.63 & 1.86 & 10.25 & 45 & -0.57 ± 0.01 & -1.76 ± 0.03 & -8.02 ± 0.72 & 0.04 ± 0.01 & 27 & 73 & Intermediate \\
Abell 2537 & 347.0922 & -2.1928 & 0.297 & 19.21 & 1.98 & 12.48 & 69 & -0.51 ± 0.01 & -2.79 ± 0.03 & -7.42 ± 0.05 & -0.26 ± 0.01 & 11 & 23 & Relaxed \\
Abell 2631 & 354.4076 & 0.2671 & 0.273 & 18.99 & 1.60 & 6.56& 90 & -0.88 ± 0.03 & -1.68 ± 0.02 & -6.46 ± 0.17 & 0.87 ± 0.01 & 138 & 34 & Disturbed \\
ZwCl2341.1+0000 & 355.8981 & 0.3313 & 0.270 & 18.96 & 1.78 & 8.94 & 42 & -1.25 ± 0.04 & -0.75 ± 0.01 & -6.74 ± 0.40 & 1.50 ± 0.01 & 9 & 322 & Disturbed \\
\end{longtable}
\end{scriptsize}
\end{landscape}


\section{Radar charts of the dynamical state proxies}

\begin{figure*}[h!]
    \centering
    \includegraphics[width=1\textwidth]{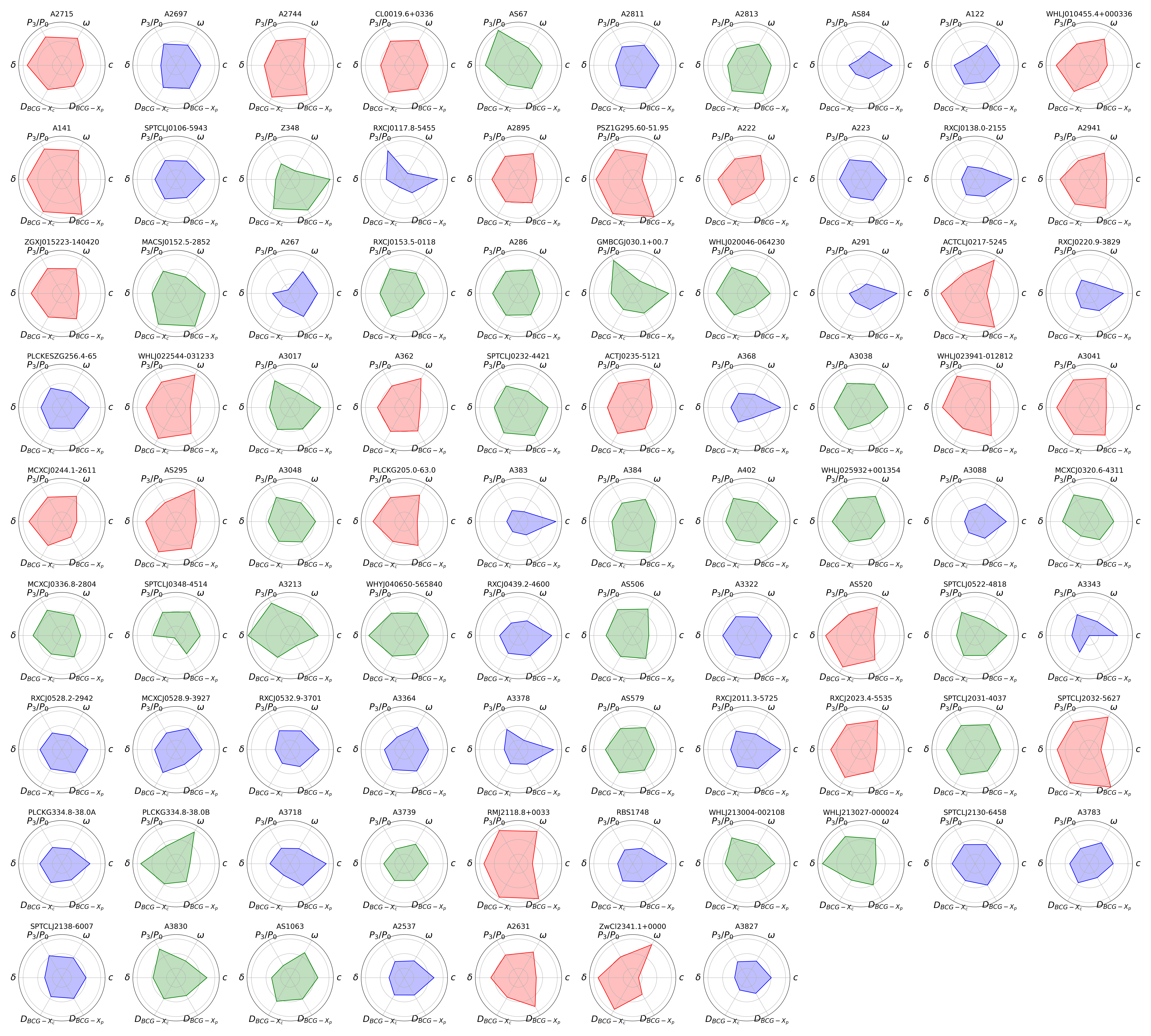}
    \caption{Radar charts with the six dynamical parameters used in this work for all clusters in the sample. The colors blue, green, and red indicate the relaxed, intermediate, and disturbed dynamical states, respectively. The parameter values are scaled in the same way as in Fig. \ref{fig:median-values-dynamical-parameters}.}
    \label{fig:radar-charts}
\end{figure*}

\end{appendix}

\end{document}